\documentclass{aa}  

\usepackage{graphicx}
\usepackage{txfonts}
\usepackage{lipsum}
\usepackage{subcaption}         
\usepackage{lscape}            
\usepackage{placeins}          

\begin{document}

   \title{Revisiting candidates for non-pulsating stars located in the Cepheid
          instability strip in the Large Magellanic Cloud} 

   \subtitle{Photometric and spectroscopic analysis} 
   \titlerunning{Revisiting candidates for non-pulsating stars located in the
                 Cepheid IS in the LMC}

   \author{W. Narloch\inst{1}
        \and C. Ga\l an\inst{1} 
        \and G. Pietrzy\'nski\inst{1} 
        \and B. Pilecki\inst{1}
        \and W. Gieren\inst{2}
        \and R. Smolec\inst{1} 
        \and H. Netzel\inst{1} 
        \and P. Wielg\'orski\inst{1}
        }

   \institute{Nicolaus Copernicus Astronomical Center, Polish Academy of Sciences, 
             Bartycka 18, 00-716 Warszawa, Poland\\
             \email{wnarloch@camk.edu.pl}
%              \thanks{Based on data from the Las Campanas Observatory.}
            \and Universidad de Concepción, Departamento de Astronomia, Casilla 160-C, 
             Concepción, Chile\\ }

   \date{Accepted: August 15, 2026}

  \abstract
   {The instability strip (IS) is a~region on the Hertzsprung-Russell diagram occupied 
   by various types of pulsating stars. However, among them there are also stars that 
   appear photometrically stable at a~certain level of detectability. Although several 
   scenarios have been proposed to explain this phenomenon, their nature remains unknown.}
   {We analyzed photometric and spectroscopic data for $11$ candidates for non-pulsating 
   stars located in the Cepheid IS of the Large Magellanic Cloud (LMC) in order
   to investigate the reasons for the lack of pulsations.} 
   {We used available temperature calibrations based on photometric colors to estimate 
   the effective temperatures of the candidates, which served as initial parameters 
   for the spectroscopic analysis. We also applied surface brightness--color relations
   calibrated for Cepheid variables, giants, and supergiants to estimate stellar radii.
   The spectral analysis was performed using the spectral synthesis method to determine 
   the atmospheric parameters; namely, the effective temperature, metallicity, surface 
   gravity, microturbulent velocity, projected rotational velocity, and chemical 
   abundances for up to $\sim$30 elements.}
   {For most objects, only a~single spectrum was available. However, no significant
   variations in radial velocities were detected among the stars with repeat observations;
   therefore, all stars were treated as single object in the analysis. Two stars exhibit
   broad spectral lines, which may indicate high rotational velocities or possible binarity;
   however, additional spectra are required to confirm this interpretation. One of these
   objects also shows asymmetric line profiles, which might be related to the presence
   of non-radial pulsation modes causing line-shape variations. A~common feature among
   all analyzed candidates is an enhancement of barium-peak $s$-process elements compared 
   to solar values. This may indicate past mass transfer from a~companion during its 
   post-asymptotic giant branch phase.}
   {This study provides an insight into the physical parameters of candidates for 
   non-pulsating stars residing within the Cepheid IS in the LMC.
   Most of the stars have parameters very similar to Cepheids and the lack of pulsations
   remains a~mystery and challenges pulsation theory. Additional spectroscopic
   observations are required to confirm and further investigate the possible origins 
   of the observed properties of these objects.}


   \maketitle 
   \nolinenumbers

%%%%%%%%%%%%%%%%%%%%%%%%%%%%%%%%%%%%%%%%%%%%%%%%%%%%%%%%%%%%%%
\section{Introduction} 
\label{sec:intro}

The classical instability strip (IS) is defined as a~region on the Hertzsprung-Russell
(HR) diagram occupied by pulsating stars of different classes 
\citep{Cox1974,GautschySaio1996}, commonly referred to as classical pulsators. 
However, many studies indicate that a~significant fraction of the stars 
residing in this region are photometrically stable at the level of tens of 
milimagnitudes 
\citep[e.g.,][]{FernieHube1971,Butler1998,Guzik2013,Guzik2015,Murphy2015,
Narloch2019,CruzReyes2024}.  

In \citet[][hereafter Paper~I]{Narloch2019}, we selected $19$ candidates 
for non-pulsating stars located in the Cepheid IS in the Large Magellanic
Cloud (LMC) based on combined photometry from the Optical Gravitational 
Lensing Experiment \citep[OGLE,][]{Udalski2000} and Str\"omgren observations.
These stars accounted for $21 - 30 \%$ of LMC giants located in the Cepheid IS,
indicating that they constitute a~significant fraction of all giants in this
region. Nevertheless, the nature of these objects and the reasons for the lack
of pulsations remain unknown.

Several scenarios can be considered. One possibility is that some (or all) 
of these objects are unresolved binary systems, which could affect the 
observed magnitudes and colors and shift the stars on the HR diagram into 
the IS \citep[e.g.,][]{Murphy2015}. 
However, it is also worth noting that binarity itself would not necessarily
remove a~star from the region of the IS, as was shown, for example, by \cite{Pilecki2018},
where a~non-pulsating companion of a~Cepheid remains well within IS borders.
Another possibility is the presence of non-radial pulsation modes with a~large
$l$~number, occurring at different frequencies \citep[e.g.,][]{Smolec2023,Netzel2024}.
Due to geometric cancellation, the associated variability may fall below the
detection threshold, but it could manifest itself in the broadening of spectral
lines.
According to several theoretical studies, radial strange modes with a~large
$n$~number (high order radial overtones) can also occur in a~star
\citep[e.g.,][]{Buchler1997,BuchlerKollath2001,TarczayNehez2026}.
They are characterized by very small amplitudes that may not produce detectable 
variations in the light curves. 
The strange modes are excited in the uppermost layers of a~star, separated from
the stellar interior, and do not coexist with the fundamental and low-order overtone
modes. They are unstable on the hot side of the blue edge of the classical IS
\citep[e.g.,][]{TarczayNehez2026}, but a~possible overlap might exist 
\citep[e.g.,][]{BuchlerKollath2001}.
Another possibility is that the apparent discrepancy arises from inaccuracies
in the derived atmospheric parameters: revised effective temperatures and 
surface gravities could move these stars outside the IS.

In this work, we performed photometric and spectroscopic analysis for the $11$ 
brightest objects from the original sample of $19$ candidates for non-pulsating
stars from the LMC Cepheid IS to address the important question regarding the
nature of these objects; namely, why they appear to occupy the IS despite the
lack of observed pulsations.
By determining their physical parameters, we aim to investigate some of the 
aforementioned scenarios or possibly identify new explanations.

The paper is organized as follows. In Sect.~\ref{sec:data} we describe 
the photometric and spectroscopic data. In Sect.~\ref{sec:analysis} we 
present the methods used and the analysis of the data. Section~\ref{sec:results} 
presents the results and discussion, and finally Sect.~\ref{sec:conclusions} 
provides a~brief summary. 

%%%%%%%%%%%%%%%%%%%%%%%%%%%%%%%%%%%%%%%%%%%%%%%%%%%%%%%%%%%%%% 
\section{Data} 
\label{sec:data} 

%-------------------------------------------------------------
\subsection{Photometry}
\label{ssec:data_phot}

Photometric data for four fields in the LMC in the Str\"omgren $uvby$ filters 
were collected using the $4.1$-m Southern Astrophysical Research (SOAR) 
telescope on Cerro Pach\'on in Chile, equipped with the SOAR Optical Imager (SOI). 
The instrument, data acquisition, and analysis are described in detail in 
Paper~I. In that work, $19$ candidates for non-pulsating stars residing in
the Cepheid IS were selected. The $V$~magnitudes, $(b-y)$ colors, and $m1$ and
$c1$ indices were obtained for these objects and are listed in
Table~3 of Paper~I. In this work, we analyze the $11$ brightest stars from the
original sample, for which we were able to obtain spectroscopic observations
(see the following Sect.~\ref{ssec:data_spec}).

Magnitudes $V$ and $I$ from the OGLE survey were obtained from the OGLE-II $BVI$
photometric maps of the LMC \citep{Udalski2000}, the OGLE-III LMC photometric 
maps \citep{Udalski2008}, and the OGLE-III LMC shallow survey \citep{Ulaczyk2012}.
Near-infrared photometry for the $11$ candidates analyzed in this paper was 
acquired in the $K_{s}$~band from the IRSF Magellanic Clouds Point Source Catalog 
\citep{Kato2007} and the $K_s$~band from the VISTA Magellanic Clouds (VMC) survey
\citep{Cioni2011}.
The photometric properties of the analyzed candidates are summarized in
Table~\ref{tab:phot_sum}.

%-------------------------------------------------------------
\subsection{Spectroscopy}
\label{ssec:data_spec}

Eleven objects from the sample of candidates for non-pulsating stars in the
Cepheid IS from Paper~I were observed with the Magellan Inamori Kyocera Echelle 
(MIKE) spectrograph on the 6.5-m Magellan Clay telescope at the Las 
Campanas Observatory in Chile in December 2019, February 2020, and December 
2021\footnote{Observations were collected within observing programs: CN2019B-74, 
PI: W. Narloch; CN2020A-98, PI: W. Gieren; CN2021B-44, PI: W. Narloch.}.
A~$5 \times 0.7$~arcsec slit was used, providing a~spectral resolution of 
$R \sim 42{,}000$ and $\sim 32{,}000$ in the blue ($\sim 3350$--$5050$\,\AA) 
and red ($\sim 4860$--$9390$\,\AA) arms, respectively. 
The journal of spectroscopic observations is presented in Table~\ref{tab:obslog}. 
Four objects were observed more than once.

Radial velocities were measured from the red-arm spectra using the {\sl fxcor} 
task of the IRAF\footnote{IRAF is distributed by the National Optical 
Astronomy Observatories, which are operated by the Association of Universities 
for Research in Astronomy, Inc., under cooperative agreement with the National 
Science Foundation.} package via cross-correlation with a~synthetic spectrum 
($T_{\rm eff} = 6000$\,K, $\log g = +1.5$\,dex, and $\rm {[M/H]} = -0.25$\,dex) from 
the POLLUX\footnote{POLLUX Database of Stellar Spectra: http://pollux.oreme.org} 
database \citep{Palacios2010}. For the cross-check broadening function (BF)
implemented in the RaveSpan code \citep{Pilecki2017} was also used resulting
in very similar values. 
Only a~single spectrum was available for most objects; however, for objects
with multiple spectroscopic observations, no significant radial-velocity variations 
were detected within the uncertainties. Therefore, all objects were treated as 
single stars in the analysis.

All spectra were corrected to the heliocentric frame and shifted according to
the measured stellar radial velocities. 
When multiple observations existed, they were averaged to produce a~single 
spectrum with an improved signal-to-noise ratio (S/N). 
As is illustrated in Fig.~\ref{fig:multi_spec},
there are no significant changes in the line profiles of the individual spectra beyond 
the level expected from the noise, justifying them being combined into a~single 
spectrum.
The final spectra used for the analysis had a~S/N ranging from $\sim 40$ to about
$95$.

Several objects exhibit very broad spectral lines, which may indicate that these
stars have high rotational velocities.
The profiles of the cross-correlation and BF functions for two objects, CAN-08
and CAN-16, exhibit clear asymmetries, which are particularly pronounced for CAN-16.
This may suggest that these objects have a~binary nature. However, spectra are
available from only a~single night for each star, and additional observations would
be required to confirm binarity.

%%%%%%%%%%%%%%%%%%%%%%%%%%%%%%%%%%%%%%%%%%%%%%%%%%%%%%%%%%%%%% 
\section{Methods and analysis} 
\label{sec:analysis} 

%-------------------------------------------------------------
\subsection{Photometry}
\label{ssec:analysis_phot}

The Str\"omgren photometric system is very useful for the classification 
of stars, as shown in Paper~I. The $(b-y)$ color is a~good estimator of
stellar temperature and can therefore be utilized to determine the effective
temperature ($T_{\rm eff}$). Another color often used for that purpose is 
$(V-K)$. 
In addition, the metallicity of red giants in the field can be estimated using 
the $m1$ index and a~metallicity calibration; for example, that given by 
\citet{Hilker2000}. 
Although this calibration is not appropriate for stars with colors similar
to Cepheids, it can provide an approximate estimate of the metallicity based
on the red giants in their surroundings. In Table~\ref{tab:phot_sum}, we provide
the average metallicity of all red giants in the field of a~given candidate,
as well as within a~radius of about $50$~arcsec (corresponding to $350$~pixels
of the SOI camera) around it.

Based on the combination of optical and near-infrared photometry, stellar
angular diameters ($\phi$) can be estimated using an appropriate surface 
brightness--color relation (SBCR). By adopting a~distance to the LMC, the 
corresponding stellar radii ($R$) can then be derived.
The stellar parameters obtained in this way can then be used as inputs for  
the spectroscopic analysis. 

\subsubsection{Correcting for reddening}
\label{ssec:redd}

The first step in the photometric analysis is the correction of the magnitudes
and colors of our candidates for interstellar reddening. We obtained reddening
values for our target stars from three different LMC reddening maps: \citet{Gorski2020},
\citet{Skowron2021}, and \citet{Netzel2026}. 
The maps of \citet{Gorski2020} and \citet{Skowron2021} were both constructed 
using the color of red clump giant stars, where the latter work provides reddening
in the form of $E(V-I)$, which we transformed to $E(B-V)$ by dividing it by $1.318$
\citep{Skowron2021}.
\citet{Netzel2026} estimated the reddening using the spectral energy distribution 
(SED) fitting technique. 
The reddening values for the analyzed candidates are listed in Table~\ref{tab:phot_sum}. 
In the calculations presented in this paper, we adopted the average reddening
value derived from the three maps.

To calculate total extinction and the ratio of total to selective extinction for
the bands summarized in Table~\ref{tab:phot_sum}, we used the Python package
\textit{extinction}\footnote{\url{https://github.com/sncosmo/extinction}}
\citep{extinction2016}, which provides implementations of commonly used extinction 
laws.
The total extinction in the $V$, $I$, and $K$ bands, as well as the corresponding 
total-to-selective extinction ratios, were obtained using the reddening law of 
\citet{Cardelli1989}, assuming $R_V=3.1$ and $A_V = 1.0$~mag.
To correct the Str\"omgren photometry for reddening, we adopted the same reddening 
vectors as in Paper~I for consistency. These vectors are based on \citet{SFD1998}: 
$E(b-y) = 0.772 \cdot E(B-V)$, $E(m1) = -0.269 \cdot E(b-y)$, and 
$E(c1) = 0.176 \cdot E(B-V)$. 

\subsubsection{Effective temperature from the photometry} 
\label{ssec:t_eff}

To estimate the effective temperature ($T_{\rm eff}$) of our candidates, which 
could serve as an input for theoretical models, we used the calibration of $T_{\rm eff}$ 
as a~function of [Fe/H] and color for FGK-type giant stars (luminosity class III) 
given, for example, by \citet{Alonso1999}.
The general expression is as follows: 

\begin{equation}
 \theta_{\rm eff} = a_0 + a_1X + a_2X^2 -a_3X\rm [Fe/H] + a_4\rm [Fe/H] + a_5\rm [Fe/H]^2,  
\end{equation}
 
\noindent where $ T_{\rm eff} = 5040/\theta_{\rm eff} $, $X$ represents the 
color, and $a_i$ ($i = 1, ..., 5$) are the coefficients of the fit. For specific 
color ranges, the coefficients are given in Table~2 in \citet{Alonso1999}.

The reported mean variation, $\Delta T_{\rm eff} / \Delta (b-y)$, amounts to about 
$62$~K per $0.01$~mag for $(b-y) < 0.45$~mag and $26$~K per $0.01$~mag for 
$(b-y) > 0.45$~mag. 
Thus, when using this calibration, an error of $0.02$~mag in $(b-y)$ implies mean 
errors of about $0.8\% - 1.5\%$ in temperature.
An uncertainty of $0.5$~dex in [Fe/H] results in a~mean temperature error 
of about $0.5\% - 1.9\%$. For the propagation of error, we adopted an uncertainty 
for [Fe/H] of $0.3$~dex, which is a~typical value obtained with the Str\"omgren method 
of metallicity estimation for red giants. 

The coefficients from the \citet{Alonso1999} relations were later updated by
\citet[][see Table~3 therein]{RamirezMelendez2005}.
In their calibration, $T_{\rm eff}$ is additionally corrected by a~polynomial 
of the form $P(X, \rm [Fe/H]) = \Sigma_i P_i X^i$, so that  
$T_{\rm eff} = 5040/\theta_{\rm eff} + P(X, \rm [Fe/H])$. 
The reported standard deviations of this calibration are $68$~K for $(b-y)$ and 
$28$~K for $(V-K)$. The derived temperatures and their uncertainties, obtained 
through error propagation including the photometric and metallicity uncertainties, 
are given in Table~\ref{tab:par_phot}.

For the calculation based on the $(V-K)$ color, we used the $V$~magnitude from 
the OGLE-III catalog and the $K_s$~magnitude from the VMC survey (see 
Table~\ref{tab:phot_sum}), converted to the 2MASS photometric system using 
equations from \citet{GonzalezFernandez2018}. 
The applied reddening vectors were calculated as described in Sect.~\ref{ssec:redd}, 
resulting in $A_{V,OGLE} = 1.038 \cdot A_V$, $A_{K,VMC} = 0.118 \cdot A_V$,
and $E(V-K) = 2.853 \cdot E(B-V)$.
The final $T_{\rm eff}$ values of the stars, together with their uncertainties 
obtained through error propagation, are reported in Table~\ref{tab:par_phot}. 
The candidate CAN-09 appears to be mismatched in the OGLE-III catalog; therefore, 
for this star only, we adopted the $V$~magnitude from the OGLE-III shallow survey, 
which we marked with an asterisk in Table~\ref{tab:par_phot}.

\subsubsection{Angular diameters and radii of stars} 
\label{ssec:phi_R_M}

The combination of $V$ and $K$~magnitudes allows the angular diameters and radii 
of stars to be derived once an appropriate SBCR is applied. Different SBCRs are
calibrated for specific types of stars, and in the case of our candidates, we are
dealing with Cepheid-like objects, which are giants and supergiants. Therefore,
we adopted SBCRs calibrated specifically for Cepheids, since our stars reside in the
IS and are expected to have similar spectral types. For the calculations, we used
$V$~magnitudes from the OGLE-III catalog and $K_s$~magnitudes from the VMC survey, 
converted to the 2MASS photometric system and de-reddened as described in 
Sect.~\ref{ssec:redd}. 

The angular diameter ($\phi$) of a~star can be expressed using the definition of
the surface brightness: 

\begin{equation}
 \phi\,{\rm (mas)} = 10^{0.2 \cdot (S - m_0)} 
 \label{eq:phi}
,\end{equation}

\noindent where $S$ is a~surface brightness in a~given band (e.g., the $V$~band) 
and $m_0$ is the de-reddened magnitude of the star in that passband. 
By definition, the angular diameter of a~star (expressed in radians) is equal 
to $\phi\,\rm (rad) = 2R / d$, where $R$ is the radius of a~star and $d$ is the 
distance. This equation yields the expression for the stellar radius:

\begin{equation}
 R\,{\rm (R_{\odot})} = \frac{\phi\,{\rm (mas) \cdot d\,{\rm (pc)}}}{9.301}
,\end{equation}

\noindent assuming for the conversion the nominal solar radius to be
$R_{\odot} = 695\,700$~km and $1 \rm pc = 3.08567750 \cdot 10^{13}$~km
\citep{Mamajek2015}.
The distance to the LMC can be adopted from \citet{Pietrzynski2019}.

We calculated the surface brightness of our candidates using several calibrations. 
First, we applied the relation of \citet{Kervella2004} calibrated for Cepheids, 
where the $K$~band is given in the South African Astronomical Observatory (SAAO) 
photometric system. Our VMC $K_s$~band magnitudes expressed in the 2MASS photometric
system were transformed to the SAAO system using the transformation equations 
from \citet{Koen2007}. 
We also used the SBCR calibration for Cepheids from \citet{Bailleul2025}, where 
the $K$~band corresponds to the Cerro Tololo Inter-American Observatory (CTIO) 
A~Novel Dual Imaging CAMera (ANDICAM) system. Our VMC $K_s$~band magnitudes
expressed in the 2MASS photometric system were transformed to the CTIO/CIT-like
(California Institute of Technology) system using the transformation equations
from \citet{Carpenter2001}.
Finally, we applied SBCR calibrations for Cepheid, as well as for stable giants 
and supergiants from \citet{FouqueGieren1997}. In this calibration, the 
$K$~magnitude is given in the Johnson photometric system; therefore, we again 
used the transformation equations from \citet{Carpenter2001} to convert our VMC 
$K_s$~band expressed in the 2MASS magnitudes.
The resulting stellar radii are listed in Table~\ref{tab:par_phot}, together 
with their uncertainties derived through error propagation (including magnitude
and reddening uncertainties, distance uncertainty, as well as photometric 
transformation and SBCR calibration uncertainties).

%-------------------------------------------------------------
\begin{table*}[ht!] 
\small
\caption{\label{tab:par_phot} Stellar parameters derived using 
photometry.}
\centering
\begin{tabular}{lccccccccccc}
\hline\hline 
CAN
& 01        & 03        & 07        & 08        & 09        & 10        & 11        & 13        & 14        & 16        & 17        \\
\hline 
\multicolumn{12}{l}{calibration from \citet{Alonso1999} for $(b-y)$, 
$E(b-y) = 0.772 \cdot E(B-V)$:} \\ 
$T_{\mathrm{eff}}$                                                        
& 5839      & 6900        & 5910      & 6768      & 5647      & 6552      & 6377      & 5744      & 5256      & 6558        & 5907      \\ 
$\sigma_{T_{\mathrm{eff}}}$                                                        
& $\pm$78   & $\pm$100    & $\pm$82   & $\pm$90   & $\pm$81   & $\pm$80   & $\pm$88   & $\pm$85        & $\pm$69   & $\pm$81     & $\pm$76   \\ 
\multicolumn{12}{l}{calibration from \citet{Alonso1999} for $(V-K)$, 
$E(V-K) = 2.855 \cdot E(B-V)$:} \\ 
$T_{\mathrm{eff}}$                                                        
& 5785      & 6271        & 5634      & 6330      & 5606$^{*}$ & 6211      & 5690      & 5720      & 5407      & 5730        & 5745      \\ 
$\sigma_{T_{\mathrm{eff}}}$                                                        
& $\pm$76   & $\pm$49     & $\pm$41   & $\pm$52   & $\pm$63    & $\pm$49   & $\pm$41   & $\pm$42       & $\pm$38   & $\pm$44      & $\pm$42   \\ 
\multicolumn{12}{l}{calibration from \citet{RamirezMelendez2005} for $(b-y)$, 
$E(b-y) = 0.772 \cdot E(B-V)$:} \\ 
$T_{\mathrm{eff}}$                                                        
& 5781      & 6927        & 5847      & 6678     & 5608      & 6466      & 6333      & 5705      & 5255      & 6446        & 5850      \\
$\sigma_{T_{\mathrm{eff}}}$                                                        
& $\pm$93   & $\pm$178    & $\pm$99   & $\pm$153 & $\pm$86   & $\pm$133  & $\pm$131  & $\pm$93        & $\pm$61   & $\pm$132    & $\pm$95   \\ 
\multicolumn{12}{l}{calibration from \citet{RamirezMelendez2005} for $(V-K)$, 
$E(V-K) = 2.855 \cdot E(B-V)$:} \\ 
$T_{\mathrm{eff}}$                                                        
& 5769      & 6382       & 5602      & 6425     & 5572$^{*}$ & 6276      & 5672      & 5703      & 5358      & 5705       & 5726      \\
$\sigma_{T_{\mathrm{eff}}}$                                                        
& $\pm$88   & $\pm$67    & $\pm$48   & $\pm$68  & $\pm$72    & $\pm$63   & $\pm$49   & $\pm$50         & $\pm$41   & $\pm$51    & $\pm$49   \\ 
\hline
\multicolumn{12}{l}{\citet{Kervella2004} for Cepheids:} \\ 
$R$                                                        
& 34.74     & 36.80      & 28.65     & 23.40    & 26.96$^{*}$  & 42.05   & 45.01    & 28.42       & 60.30     & 25.53      & 51.82    \\
$\sigma_R$                                                        
& $\pm$1.34  & $\pm$1.03    & $\pm$0.82  & $\pm$0.66  & $\pm$1.06  & $\pm$1.18  & $\pm$1.27 & $\pm$0.80     & $\pm$1.71    & $\pm$0.74  & $\pm$1.45    \\ 
\multicolumn{12}{l}{\citet{Bailleul2025} for Cepheids:} \\ 
$R$                                                        
& 33.78     & 35.79      & 27.86     & 22.75    & 26.22$^{*}$  & 40.89   & 43.77    & 27.64      & 58.63     & 24.82      & 50.39    \\
$\sigma_R$                                                        
& $\pm$1.33  & $\pm$1.03    & $\pm$0.82  & $\pm$0.66  & $\pm$1.05  & $\pm$1.18  & $\pm$1.27 & $\pm$0.81     & $\pm$1.71    & $\pm$0.74  & $\pm$1.46    \\ 
\multicolumn{12}{l}{\citet{FouqueGieren1997} for Cepheids:} \\ 
$R$                                                        
& 34.63     & 36.77      & 28.57     & 23.38    & 26.86$^{*}$  & 42.00   & 44.88    & 28.34       & 60.46     & 25.43      & 51.61    \\
$\sigma_R$                                                        
& $\pm$4.21  & $\pm$4.24    & $\pm$3.44  & $\pm$2.69  & $\pm$3.32  & $\pm$4.86  & $\pm$5.37  & $\pm$3.39     & $\pm$7.37    & $\pm$3.04  & $\pm$6.15  \\ 
\multicolumn{12}{l}{\citet{FouqueGieren1997} for giants:} \\ 
$R$                                                        
& 35.75     & 38.34      & 29.38     & 24.40    & 27.61$^{*}$  & 43.74   & 46.22    & 29.21       & 61.42     & 26.22      & 53.22    \\
$\sigma_R$                                                        
& $\pm$2.54  & $\pm$2.49    & $\pm$1.99  & $\pm$1.58  & $\pm$2.01  & $\pm$2.84  & $\pm$3.11 & $\pm$1.96     & $\pm$4.24    & $\pm$1.77  & $\pm$3.56   \\ 
\multicolumn{12}{l}{\citet{FouqueGieren1997} for supergiants:} \\ 
$R$                                                        
& 37.22     & 40.21      & 30.52     & 25.61    & 28.66$^{*}$  & 45.83   & 48.05    & 30.38      & 63.54     & 27.27      & 55.38    \\
$\sigma_R$                                                        
& $\pm$4.50  & $\pm$4.62    & $\pm$3.66  & $\pm$2.94  & $\pm$3.52  & $\pm$5.29  & $\pm$5.74 & $\pm$3.62     & $\pm$7.78    & $\pm$3.25  & $\pm$6.58  \\ 
\hline
\multicolumn{12}{l}{Radii calculated using $M_{bol}$ and $T_{\rm eff}$ from spectroscopy:} \\ 
$R$                                                        
& 34.46     & 64.59      & 27.07     & 43.53    & 26.52$^{*}$  & 68.66   & 54.53    & 27.78      & 47.46     & 34.14      & 54.24    \\
$\sigma_R$                                                        
& $\pm$1.72  & $\pm$2.20    & $\pm$1.56  & $\pm$2.68  & $\pm$1.38  & $\pm$1.97  & $\pm$1.74 & $\pm$1.09     & $\pm$1.65    & $\pm$1.96  & $\pm$2.70  \\ 
$BC_V^{**}$ 
& -0.11     & -0.03      & -0.11     & 0.00     & -0.09        & -0.04   & -0.04    & -0.11  & -0.15     & -0.01      & -0.08    \\ 
\hline
\end{tabular}
\tablefoot{ \\ {Units are: Kelvin (K) for $T_{\rm eff}$ and ($R_{\odot}$) for radii.} \\
\tablefoottext{*}{$V$ magnitude adopted from the OGLE-III LMC shallow survey \citep{Ulaczyk2012}.} \\ 
\tablefoottext{**}{$BC_V$ are expressed in magnitudes, calculated using calibration of
\citet{Alonso1999} and the spectroscopic values of $T_{\mathrm{eff}}$ and [M/H].} \\
}
\end{table*}

%-------------------------------------------------------------
\subsection{Spectroscopy}
\label{ssec:analysis_spec}

The spectral synthesis method was applied to determine the atmospheric
parameters: $T_{\rm eff}$, metallicity [M/H], surface gravity ($\log g$), 
microturbulent velocity ($\xi_{\rm t}$), and projected rotational velocity 
($V_{\rm rot} \sin i$), as well as abundances for up to $\sim$30 chemical 
elements (summarized in Tables~\ref{tab:chem_sum} and \ref{tab:chem_sum2}). 
The spectral synthesis approach was adopted to avoid the line-blending
problems associated with rotational broadening in relatively fast-rotating 
stars, following a procedure similar to that described by 
\citet{Graczyk2021,Graczyk2022,Graczyk2025}.
We used the {\sl single} version of the Grid Search in Stellar Parameters
(GSSP) software package \citep{Tkachenko2015}, which employs the local-thermal-equilibrium-based
radiative transfer code {\small SYNTH}V \citep{Tsymbal1996}.

The atmospheric model grid provided by default with the GSSP code -- 
the LL{\small MODELS} grid \citep{Shulyak2004}, extended by a~MARCS 
grid \citep{Gustafsson2008} for lower effective temperatures ($T_{\rm eff} <
5600$~K) -- does not include models with $\log g < 2.5$~dex, making it 
unsuitable for our sample, which contains stars with lower surface gravities. 
A~proper solution requires the simultaneous determination of $\log g$ due to its 
correlation with other parameters. Fixing $\log g$ at $2.5$~dex results 
in biased estimates of other atmospheric parameters, overestimating 
$T_{\rm eff}$ and [M/H] in particular. 

We adopted the original Kurucz\footnote{http://kurucz.harvard.edu/}
atmosphere models \citep{Kurucz1993} at their original resolution, adapted to
the format compatible with the GSSP code, covering the following parameter
ranges: $T_{\rm eff} = 3500-8750$~K, $\log g = 0.0 - 5.0$~dex, and
$\rm [M/H] = -1.0$ to $+1.0$~dex. The available grid steps for parameter changes
were: $\Delta T_{\rm{eff}} = 250$~K, $\Delta \log{g} = 0.5$~dex, and
$\Delta \rm [M/H] = 0.2$~dex in the vicinity of the value of solar metallicity.

The procedure for the spectral synthesis method was as follows. The analyzed 
spectral ranges covered about $4400-7530$~\AA. Regions around the hydrogen 
H$\alpha$ and H$\beta$ lines were excluded from the analysis, as their broad
wings hinder a~reliable placement of the continuum level and may lead to
significant distortions in the measured line depths.

Additionally, dozens of narrow spectral regions were identified and excluded 
using a~custom program that generates two types of masks to remove problematic 
or noisy areas: (I) removes noisy regions with no spectral lines deeper
than $2.5\sigma$ of the noise, while (II) eliminates features that have
no counterparts in the line lists, as well as cases in the synthetic spectra
with inaccurate atomic data.
Furthermore, regions affected by artifacts from imperfectly removed atmospheric 
lines -- water (H$_2$O: $\lambda \sim 5870-6000$~\AA\ and $7159-7378$~\AA) 
and molecular oxygen (O$_2$: $\lambda \sim 6274-6330$~\AA\ and $6860-7053$~\AA) 
-- were also excluded.

%-------------------------------------------------------------
   \begin{figure*}[ht!]
   \centering
   \includegraphics[width=0.85\hsize]{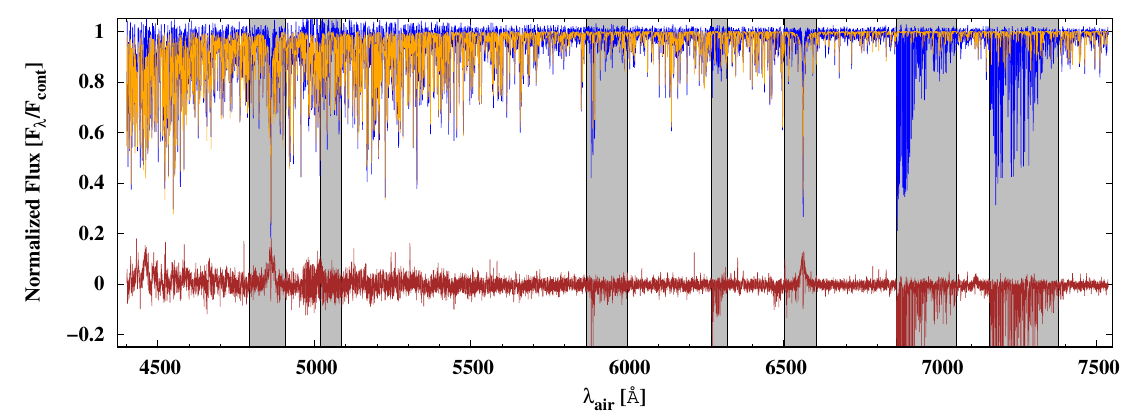}
      \caption{Comparison of the synthetic spectrum (orange) with the observed spectrum 
               of CAN-17 (blue). The residuals, calculated in the regions used for the 
               analysis, are shown below with a~brown line. Shaded gray areas, containing 
               the hydrogen H$\alpha$ and H$\beta$ lines with broad wings, as well as 
               telluric lines from H$_2$O and O$_2$ molecules, were excluded from the 
               calculations.}
         \label{fig:ex_syntsp_Can-17} 
   \end{figure*}
%-------------------------------------------------------------

Calculations began by assigning relatively coarse but wide grids of fit 
parameters to approximately locate the region near the global minimum. The 
free parameters were $T_{\rm eff}$, [M/H], $\log g$, $\xi_{\rm t}$, and 
$V_{\rm rot} \sin i$. The initial grid was centered around the parameter 
values determined in Sect.~\ref{ssec:analysis_phot}. 
Initial values of $\xi_{\rm t}$ were estimated using published correlations 
with $\log g$ and $T_{\rm eff}$ \citep{Gray2001}.

The macroturbulent velocity ($\zeta_{\rm t}$) was not treated as a~free
parameter because of its strong degeneracy with rotational velocity. Instead, 
it was estimated from published empirical relations derived from analysis of 
stellar line broadening in stars of F, G, and K spectral type 
\citep{Smalley2014,Gray2005}, kept fixed during the calculations, and adjusted 
according to changes in $T_{\rm eff}$. Consequently, the value of
$V_{\rm{rot}} \sin{i}$ derived by GSSP should be interpreted in the context 
of the adopted macroturbulent velocity, as the observed line broadening depends 
on the combined contribution of rotation and macroturbulence, approximately 
described by $\sqrt{(V_{\rm{rot}} \sin{i})^2+\zeta t^2}$.

Unlike classical Cepheids, for which line broadening is usually dominated
by macroturbulent motions associated with pulsation, the objects analysed 
here may show a~significant contribution from rotational broadening. This 
difference should be considered when comparing these Cepheid-like objects 
with classical pulsators.

Parameter ranges were progressively narrowed to allow finer sampling in
subsequent iterations with GSSP. The first step to locate the approximate
solution area was performed without additional masking. In the second step,
a~combination of masks (I) and (II) was applied to evaluate the
approximate region of best-fit atmospheric parameters in parameter space.

Next, the parameters were fixed at these best values, and the chemical 
composition for about three dozen elements was determined in the third step
using only (I) masks, and again in the fourth step using a~combination 
of (I) and (II) masks. Finally, a~fifth step was performed using full masking, 
with metallicity fixed and the newly derived chemical composition, to refine 
the values of $T_{\rm eff}$, $\log g$, $\xi_{\rm t}$, and $V_{\rm rot} \sin i$. 
Steps four and five were repeated once to derive the final parameters.

\begin{figure*}[ht!]
   \centering
   \includegraphics[width=0.85\hsize]{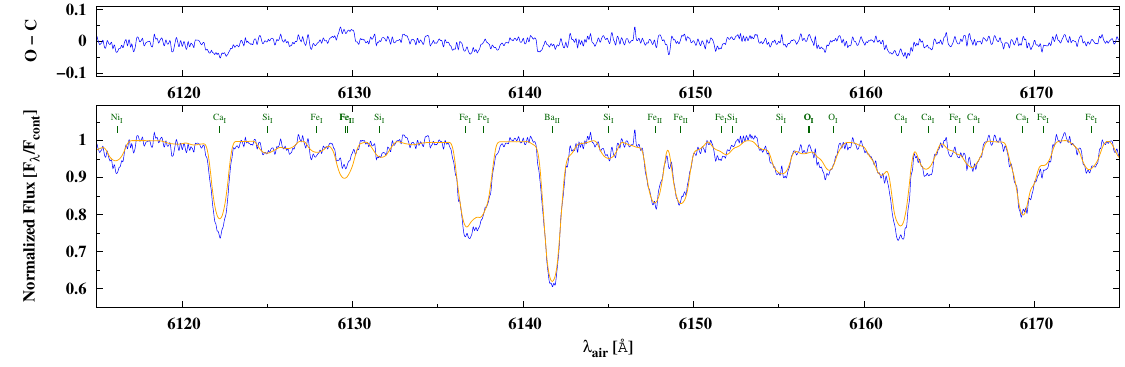}
      \caption{Zoom-in comparison of synthetic (orange) and observed (blue) spectra 
               of CAN-17 Fig.~\ref{fig:ex_syntsp_Can-17}, centered on the 
               Ba$_{\rm{II}}$ ($\lambda 6141.7$~\AA).
               Residuals indicate a~satisfactory agreement between the synthetic 
               and observed spectra.}
         \label{fig:exp_BaII_Can-17} 
   \end{figure*}

The $1\sigma$ uncertainties were estimated by finding the intersection of 
the $1\sigma$ levels in $\chi^2$ ($\chi^2_{1\sigma}$) with polynomial functions 
fit to the minimum values of the reduced $\chi^2$, as recommended by 
\citet{Tkachenko2015}. The resulting final parameters are given in Table~\ref{tab:t_res}. 
Examples of comparisons between the observed spectra of CAN-17 over the full 
analyzed range and its best-fit synthetic spectra are shown in 
Fig.~\ref{fig:ex_syntsp_Can-17}, and expanded regions around the Ba$_{\rm{II}}$
($\lambda 6141.7$~\AA) line are shown in Fig.~\ref{fig:exp_BaII_Can-17}.

The classical equivalent width (EW) method was also applied to verify the
results obtained with the GSSP code. This method relies on the excitation 
and ionization balance of iron lines, using the iron abundance as a~proxy for 
metallicity. It is reliable for most FGK-type stars and is therefore suitable 
for our sample. The same parameters were determined as in the spectral synthesis, 
except for $V_{\rm rot} \sin i$.

Automatic EW measurements were performed using the ARESv2 code \citep{Sousa2015}.
Atmospheric parameters ($T_{\rm eff}$, $\log g$, [M/H], and $\xi_{\rm t}$) were 
then derived using the MOOG code \citep[version Nov. 2019;][]{Sneden1973}. 
Abundances were determined for a~smaller set of $11$ elements due to the limited 
availability of atomic data for the spectral lines in this approach (this 
procedure failed for CAN-08, CAN-16, and CAN-17, due to lines being too broad and
blended, which prevented reliable measurements of the EW).
The solution was performed iteratively, with outlier points being rejected at
each step using a~$3 \sigma$ criterion. The resulting parameter values are consistent
with those obtained from spectral synthesis within the uncertainties, although 
the errors from the EW method are considerably larger -- typically $\sim$3--4 
times higher (see Table~\ref{tab:t_ew}).
This EW approach, using ARES $+$ MOOG, performs well for spectra with good 
S/N ($>100$) and relatively narrow lines, and should be avoided for fast 
rotators ($V_{\rm rot} \sin i \gtrsim 15$~km\,s$^{-1}$). 
Consequently, it is at the limit of applicability for the present sample. The 
results from the classical EW method are therefore treated only as a~consistency 
check for the more precise spectral synthesis results and are not considered 
in the subsequent discussion.

%-------------------------------------------------------------
%-------------------------------------------------------------

%-------------------------------------------------------------
\begin{table*}[ht!] 
\small
\caption{\label{tab:t_res} Atmospheric parameters derived from spectral
synthesis using the GSSP code based on Kurucz model atmospheres. }
\centering
\begin{tabular}{lccccccccccc}
\hline\hline 
CAN
& 01        & 03        & 07        & 08        & 09        & 10        & 11        & 13        & 14        & 16        & 17        \\
\hline 
\hline
[M/H]                                                                    
& -0.28     & -0.36     & -0.29     & 0.04      & -0.36     & -0.48     & -0.30     & -0.30     & -0.32     & -0.06     & -0.34     \\
$\sigma_{\mathrm{[M/H]}}$                                                          
& $\pm$0.09 & $\pm$0.08 & $\pm$0.12 & $\pm$0.15 & $\pm$0.10 & $\pm$0.11 & $\pm$0.07 & $\pm$0.08 & $\pm$0.06 & $\pm$0.15 & $\pm$0.11 \\
$T_{\mathrm{eff}}$                                                       
& 5670      & 6989      & 5728      & 7174      & 5909      & 6807      & 6512      & 5716      & 5451      & 6819      & 5914      \\
$\sigma_{T_{\mathrm{eff}}}$                                                
& $\pm$131  & $\pm$99   & $\pm$156  & $\pm$210  & $\pm$134  & $\pm$73   & $\pm$84   & $\pm$98   & $\pm$79   & $\pm$184  & $\pm$136  \\
$\mathrm{log}g$                                                          
& 1.06      & 1.44      & 1.56      & 2.83      & 1.83      & 1.56      & 1.48      & 1.57      & 1.03      & 2.61      & 1.02      \\
$\sigma_{\mathrm{log}g}$                                                   
& $\pm$0.18 & $\pm$0.18 & $\pm$0.30 & $\pm$0.37 & $\pm$0.28 & $\pm$0.13 & $\pm$0.18 & $\pm$0.20 & $\pm$0.14 & $\pm$0.39 & $\pm$0.19 \\
$\xi_{\rm{t}}$                                                           
& 3.99      & 3.29      & 2.58      & 3.47      & 3.88      & 4.35      & 4.22      & 4.21      & 4.09      & 3.89      & 4.03      \\
$\sigma_{\xi_{\rm{t}}}$                                                    
& $\pm$0.14 & $\pm$0.12 & $\pm$0.15 & $\pm$0.23 & $\pm$0.19 & $\pm$0.13 & $\pm$0.15 & $\pm$0.16 & $\pm$0.13 & $\pm$0.24 & $\pm$0.14 \\
$\zeta_{\rm{t}}$\tablefootmark{a}                                             
& 3.5       & 8.0       & 3.3       & 8.0       & 4.0       & 7.5       & 6.4       & 3.4       & 2.5       & 7.8       & 4.1       \\
$\sigma_{\zeta_{\rm{t}}}$\tablefootmark{b}                                         
& $\pm$0.66 & $\pm$0.53 & $\pm$0.76 & $\pm$0.97 & $\pm$0.67 & $\pm$0.43 & $\pm$0.48 & $\pm$0.53 & $\pm$0.46 & $\pm$0.87 & $\pm$0.68 \\
$V_{\mathrm{rot}}\mathrm{sin}i$                                          
& 12.02     & 6.15      & 13.32     & 41.19     & 12.53     & 23.95     & 8.32      & 13.63     & 12.23     & 32.26     & 32.25     \\
$\sigma_{V_{\mathrm{rot}}\mathrm{sin}i}$                                   
& $\pm$0.66 & $\pm$1.19 & $\pm$0.96 & $\pm$1.99 & $\pm$0.93 & $\pm$0.64 & $\pm$0.99 & $\pm$0.83 & $\pm$0.62 & $\pm$1.58 & $\pm$0.91 \\
$\sigma^{\zeta}_{V_{\mathrm{rot}}\mathrm{sin}i}$\tablefootmark{c}                 
& $\pm$0.19 & $\pm$0.69 & $\pm$0.19 & $\pm$0.19 & $\pm$0.21 & $\pm$0.14 & $\pm$0.37 & $\pm$0.13 & $\pm$0.09 & $\pm$0.21 & $\pm$0.09 \\
$\sqrt{(V_{\mathrm{rot}}\mathrm{sin}i)^{2} + \zeta_{\rm{t}}^{2}}$        
& 12.52     & 10.09     & 13.72     & 41.96     & 13.15     & 25.10     & 10.50     & 14.05     & 12.48     & 33.19     & 32.51     \\
$\sigma_{\sqrt{(V_{\mathrm{rot}}\mathrm{sin}i)^{2} + \zeta_{\rm{t}}^{2}}}$ 
& $\pm$0.63 & $\pm$0.97 & $\pm$0.94 & $\pm$1.98 & $\pm$0.90 & $\pm$0.63 & $\pm$0.92 & $\pm$0.82 & $\pm$0.61 & $\pm$1.57 & $\pm$0.91 \\
$[\alpha / \mathrm{Fe}]$                                           & 0.11      & 0.15      & 0.18
& 0.06      & 0.10      & 0.17      & 0.09      & 0.11      & 0.07      & 0.10      & 0.12      \\
$\sigma_{[\alpha / \mathrm{Fe}]}$                                   & $\pm$0.03      & $\pm$0.03      & $\pm$0.05
& $\pm$0.05      & $\pm$0.03      & $\pm$0.02      & $\pm$0.03      & $\pm$0.03      & $\pm$0.03      & $\pm$0.05      & $\pm$0.04      \\
$[\mathrm{Ba}]$                                                   & 0.32      & 0.08      & 0.38
& 1.12      & 0.57      & 0.39      & 0.59      & 0.51      & 0.42      & 0.77      & 0.56      \\
$\sigma_{[\mathrm{Ba}]}$                                            & $\pm$0.04      & $\pm$0.12      & $\pm$0.08
& $\pm$0.04      & $\pm$0.06      & $\pm$0.03      & $\pm$0.08      & $\pm$0.07      & $\pm$0.05      & $\pm$0.08      & $\pm$0.03      \\
$[\mathrm{hs} / \mathrm{Fe}]$                                     & 0.28      & 0.42      & 0.32
& 0.43      & 0.34      & 0.41      & 0.52      & 0.39      & 0.41      & 0.41      & 0.37      \\
$\sigma_{[\mathrm{hs} / \mathrm{Fe}]}$                              & $\pm$0.03      & $\pm$0.03      & $\pm$0.05
& $\pm$0.09      & $\pm$0.04      & $\pm$0.03      & $\pm$0.03      & $\pm$0.04      & $\pm$0.03      & $\pm$0.08      & $\pm$0.04      \\
$[\mathrm{ls} / \mathrm{Fe}]$                                     & -0.14     & -0.09     & -0.09
& 0.06      & -0.06     & 0.06      & 0.05      & 0.02      & 0.01      & -0.11     & -0.01     \\
$\sigma_{[\mathrm{ls} / \mathrm{Fe}]}$                            & $\pm$0.04      & $\pm$0.03      & $\pm$0.05
& $\pm$0.08      & $\pm$0.04      & $\pm$0.03      & $\pm$0.03      & $\pm$0.04      & $\pm$0.03      & $\pm$0.07      & $\pm$0.04      \\
$[\mathrm{hs} / \mathrm{ls}]$                                     & 0.42      & 0.50      & 0.41
& 0.37      & 0.41      & 0.34      & 0.48      & 0.37      & 0.40      & 0.52      & 0.38      \\
$\sigma_{[\mathrm{hs} / \mathrm{ls}]}$                              & $\pm$0.05      & $\pm$0.04      & $\pm$0.08
& $\pm$0.12      & $\pm$0.05      & $\pm$0.05      & $\pm$0.04      & $\pm$0.05      & $\pm$0.05      & $\pm$0.10      & $\pm$0.06      \\
$[\mathrm{hs}_{\mathrm L} / \mathrm{Fe}]$                         & 0.30      & 0.45      & 0.36
& 0.43      & 0.38      & 0.44      & 0.57      & 0.44      & 0.43      & 0.43      & 0.42      \\
$\sigma_{[\mathrm{hs}_{\mathrm L} / \mathrm{Fe}]}$                  & $\pm$0.04      & $\pm$0.03      & $\pm$0.06
& $\pm$0.11      & $\pm$0.04      & $\pm$0.04      & $\pm$0.03      & $\pm$0.04      & $\pm$0.03      & $\pm$0.10      & $\pm$0.05      \\
$[\mathrm{ls}_{\mathrm L} / \mathrm{Fe}]$                         & -0.15     & -0.09     & -0.11
& 0.06      & -0.08     & 0.00      & 0.05      & 0.03      & 0.01      & -0.11     & -0.04     \\
$\sigma_{[\mathrm{ls}_{\mathrm L} / \mathrm{Fe}]}$                  & $\pm$0.04      & $\pm$0.03      & $\pm$0.06
& $\pm$0.08      & $\pm$0.04      & $\pm$0.03      & $\pm$0.03      & $\pm$0.04      & $\pm$0.04      & $\pm$0.07      & $\pm$0.05      \\
$[\mathrm{hs}_{\mathrm L} / \mathrm{ls}_{\mathrm L}]$             & 0.45      & 0.54      & 0.47
& 0.37      & 0.46      & 0.44      & 0.52      & 0.41      & 0.42      & 0.54      & 0.46      \\
$\sigma_{[\mathrm{hs}_{\mathrm L} / \mathrm{ls}_{\mathrm L}]}$      & $\pm$0.06      & $\pm$0.05      & $\pm$0.08
& $\pm$0.14      & $\pm$0.06      & $\pm$0.05      & $\pm$0.04      & $\pm$0.06      & $\pm$0.05      & $\pm$0.13      & $\pm$0.07      \\
\hline
\end{tabular}
\tablefoot{ \\ 
{Units are: Kelvin (K) for $T_{\rm eff}$, km\,s$^{-1}$ for velocity parameters, and dex 
for abundances.} \\
\tablefoottext{a}{$\zeta_{\rm{t}}$ was treated as a constant parameter in the calculations, 
with the values adopted as a function of $T_{\mathrm{eff}}$, following the relation given 
by \citet[][see their Fig.~6]{Smalley2014}.} \\
\tablefoottext{b}{The uncertainty, $\sigma_{\zeta_{\rm{t}}}$, was estimated from the 
dispersion of the $\zeta_{\rm{t}} (T_{\mathrm{eff}})$ relation, yielding a $1 \sigma$ 
uncertainty of $\sigma_{\zeta_{\rm{t}} (\zeta_{\rm{t}})} = \pm 0.146$~dex. 
The uncertainty resulting from the effective temperature error was estimated as 
$\sigma_{\zeta_{\rm{t}} (T_{\mathrm{eff}})} = (f(T_{\mathrm{eff}} + \sigma_{ T_{\mathrm{eff}})} - f(T_{\mathrm{eff}} - \sigma_{T_{\mathrm{eff}}))} / 2$.
The final uncertainty, $\sigma_{\zeta_{\rm{t}}}$, was calculated as 
$\sigma_{\zeta_{\rm{t}}} = \sigma_{\zeta_{\rm{t}} (\zeta_{\rm{t}})} + \sigma_{\zeta_{\rm{t}} (T_{\mathrm{eff}})}$.} \\
\tablefoottext{c}{The estimated 1~$\sigma$ uncertainty in $V_{rot}\mathrm{sin}i$ resulting 
from the propagation of the uncertainty in $\zeta_{\rm{t}}$ into the quantity 
$\sqrt{(V_{\mathrm{rot}}\mathrm{sin}i)^{2} + \zeta_{\rm{t}}^{2}}$ determined during the 
calculation process with the GSSP code.}
}
\end{table*}

%%%%%%%%%%%%%%%%%%%%%%%%%%%%%%%%%%%%%%%%%%%%%%%%%%%%%%%%%%%%%%
\section{Results and discussion} 
\label{sec:results}

%%%%%%%%%%%%%%%%%%%%%%%%%%%%%%%%%%%%%%%%%%%%%%%%%%%%%%%%%%%%%%

\subsection{Comparison of the empirical IS} 
\label{ssec:res_IScomp}

\begin{figure*}[ht!]
   \centering
   \begin{subfigure}{0.43\textwidth}
    \centering
     \includegraphics[width=\textwidth]{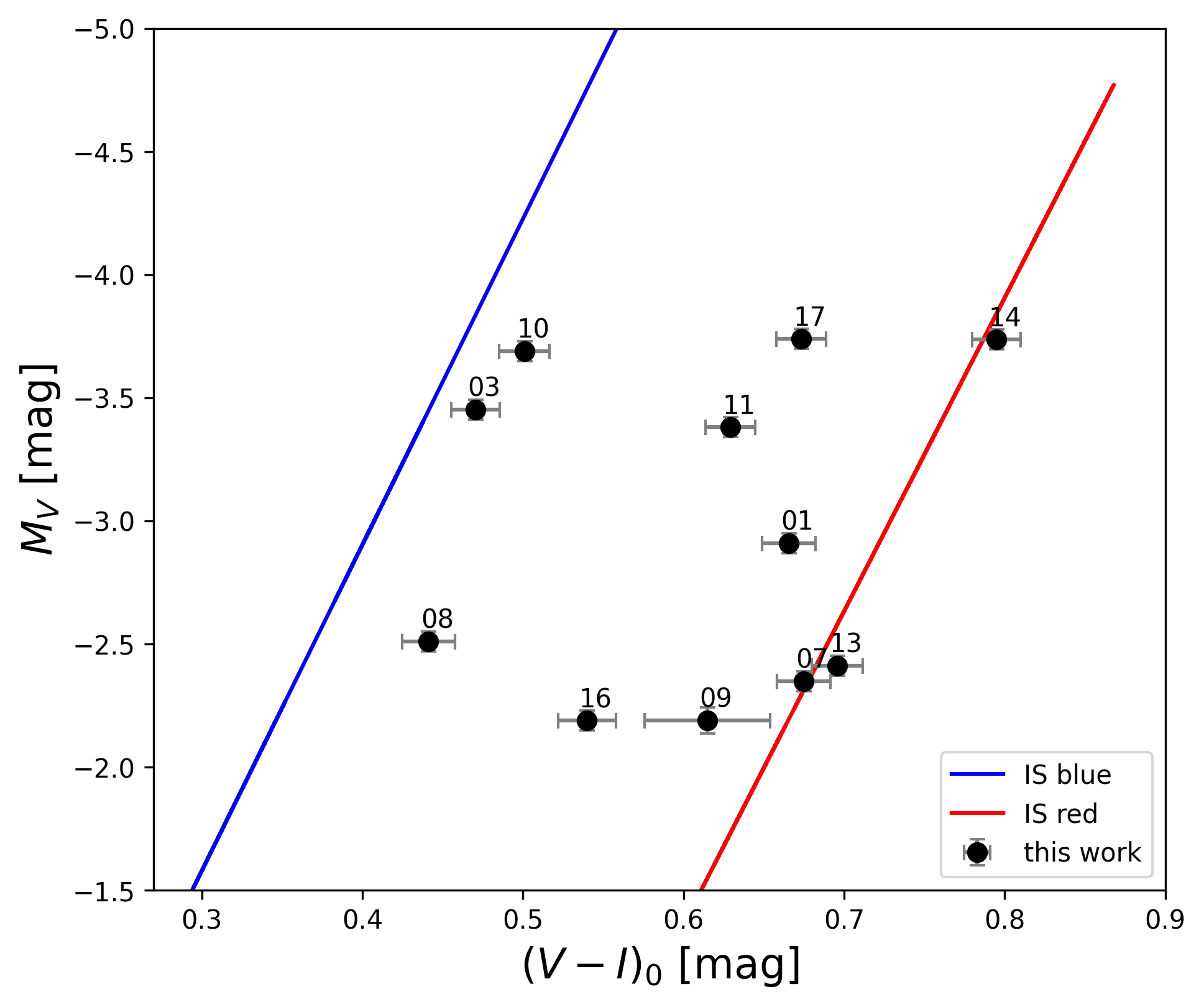}
     \label{fig:sub1}
   \end{subfigure}
   \begin{subfigure}{0.43\textwidth}
    \centering
     \includegraphics[width=\textwidth]{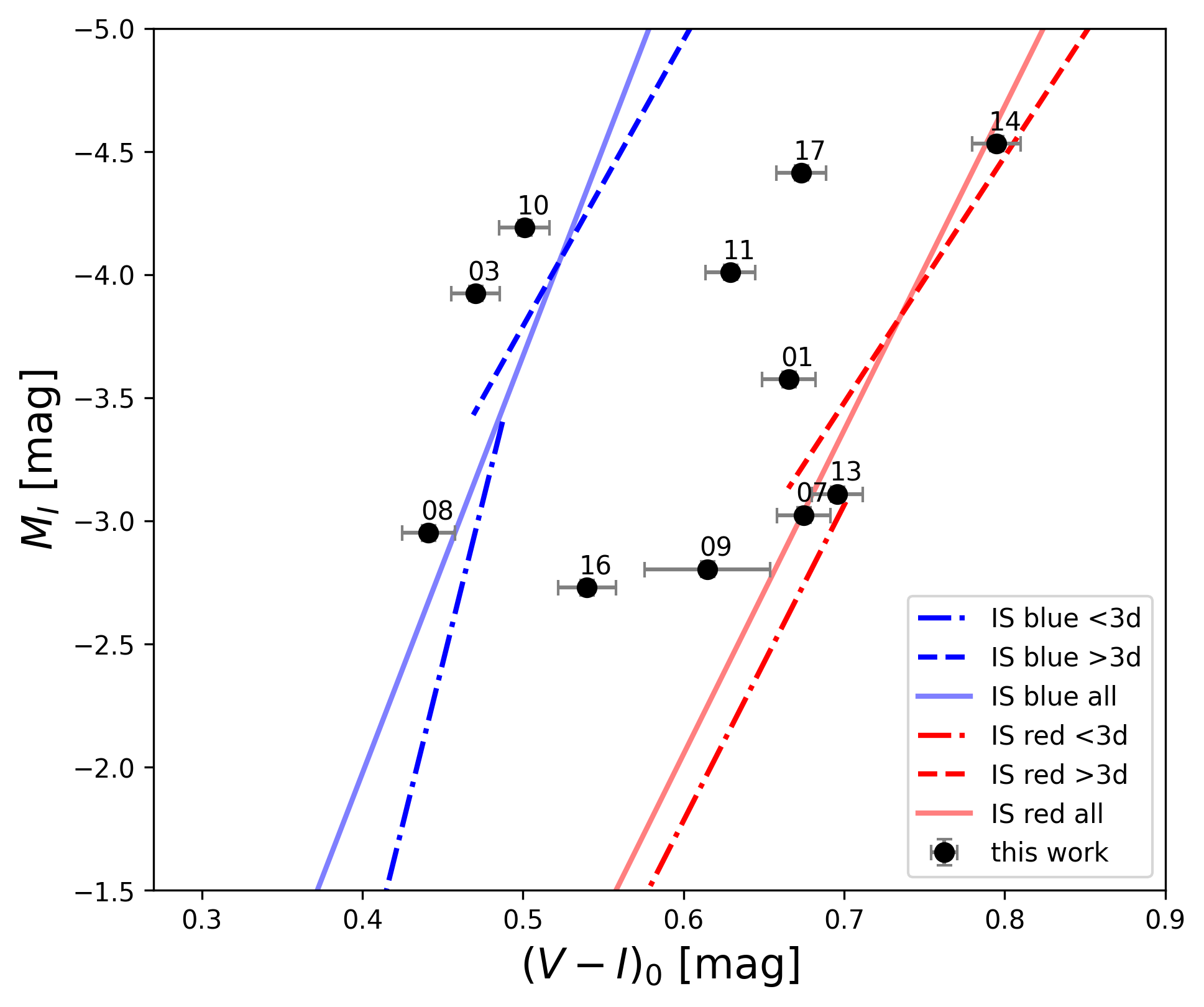}
     \label{fig:sub2}
   \end{subfigure}
   \caption{CMDs comparing empirical IS boundaries from \citet[][\textit{left}]{Narloch2019}
            and \citet[][\textit{right}]{Espinoza2024}. The number of a~given candidate is 
            marked above the points.}
   \label{fig:cmd_comp}
   \end{figure*}

Figure~\ref{fig:cmd_comp} presents the color-magnitude diagrams (CMDs)
of the $11$ candidates. 
The de-reddened $V$ and $I$ magnitudes and $(V-I)$ colors were calculated 
based on OGLE-III photometry, whereas in Paper~I, OGLE-II photometry was used. 
However, the differences are small, amounting to less than about $0.02$~mag 
in $V$ and about $0.04$~mag in $(V-I)$.
In Paper~I, the reddening values were adopted from the reddening map of 
\citet{Gorski2020}. Since then, new reddening maps have become available. 
The reddening values for our candidates were recomputed as described in 
Sect.~\ref{ssec:redd}, using three different maps. The average difference 
between the old and new values is about $0.03$~mag (see Table~\ref{tab:phot_sum}).  
The absolute magnitudes were calculated using the formula $M_V = V_0 - 5\log d + 5$, 
where $d$ is the distance to the LMC adopted from \citet{Pietrzynski2019}. 
No correction for the LMC geometry was applied, since the stars are located 
within a~relatively small central region of the galaxy and expected corrections 
are very small \citep[$\approx$$0.02$~mag;][]{Pietrzynski2019}.

In Paper~I, an empirical IS was derived for the selection of candidates.
Recently, however, a~new empirical IS was published by \citet{Espinoza2024}. 
Both ISs are shown in Fig.~\ref{fig:cmd_comp} for comparison.
The IS boundaries from Paper~I were derived for fundamental-mode (F) and
first-overtone (1O) Cepheids from the OGLE Collection of Variable Stars, 
using de-reddened magnitudes and colors based solely on the \citet{Gorski2020} 
reddening map, without applying any correction for the LMC geometry. 
Since this reddening map provides, on average, higher $E(B-V)$ values than 
the maps of \citet{Skowron2021} and \citet{Netzel2026}, two candidates in 
the left panel of Fig.~\ref{fig:cmd_comp} appear to lie slightly outside 
the red edge of the IS. 
\citet{Espinoza2024}, on the other hand, derived their IS boundaries using 
$I$~band magnitudes and $(V-I)$ colors de-reddened with the \citet{Skowron2021} 
reddening map only and corrected for the LMC geometry, resulting in 
narrower IS. 
They also determined the IS boundaries for all F and 1O Cepheids over the
full range of pulsation periods, as well as separately for Cepheids with periods
longer and shorter than $3$~days.
In the CMD shown in the right panel of Fig.~\ref{fig:cmd_comp}, three of our 
candidates appear to lie outside the IS on the blue side.

This comparison demonstrates that the position and width of the empirical
IS are sensitive to the adopted reddening values and the methodology used 
for its derivation. Therefore, any future selections of stars from the 
empirical IS region should be performed consistently with the adopted reddening 
map and calibration procedure. 
Overall, however, the comparison shows good agreement between the different 
empirical IS boundaries and supports the classification of our candidates 
as non-pulsating stars located in the Cepheid IS in the LMC, with a~maximum 
of three outliers depending on the empirical IS used. The differences resulting 
from the updated photometry and reddening values are much smaller than the 
width of the IS itself. Consequently, the conclusions from Paper~I remain 
unchanged.

%-------------------------------------------------------------
%-------------------------------------------------------------

\subsection{Comparison of stellar radii} 
\label{ssec:res_radii}

\begin{figure*}[ht!]
   \centering
   \begin{subfigure}{0.43\textwidth}
    \centering
     \includegraphics[width=\textwidth]{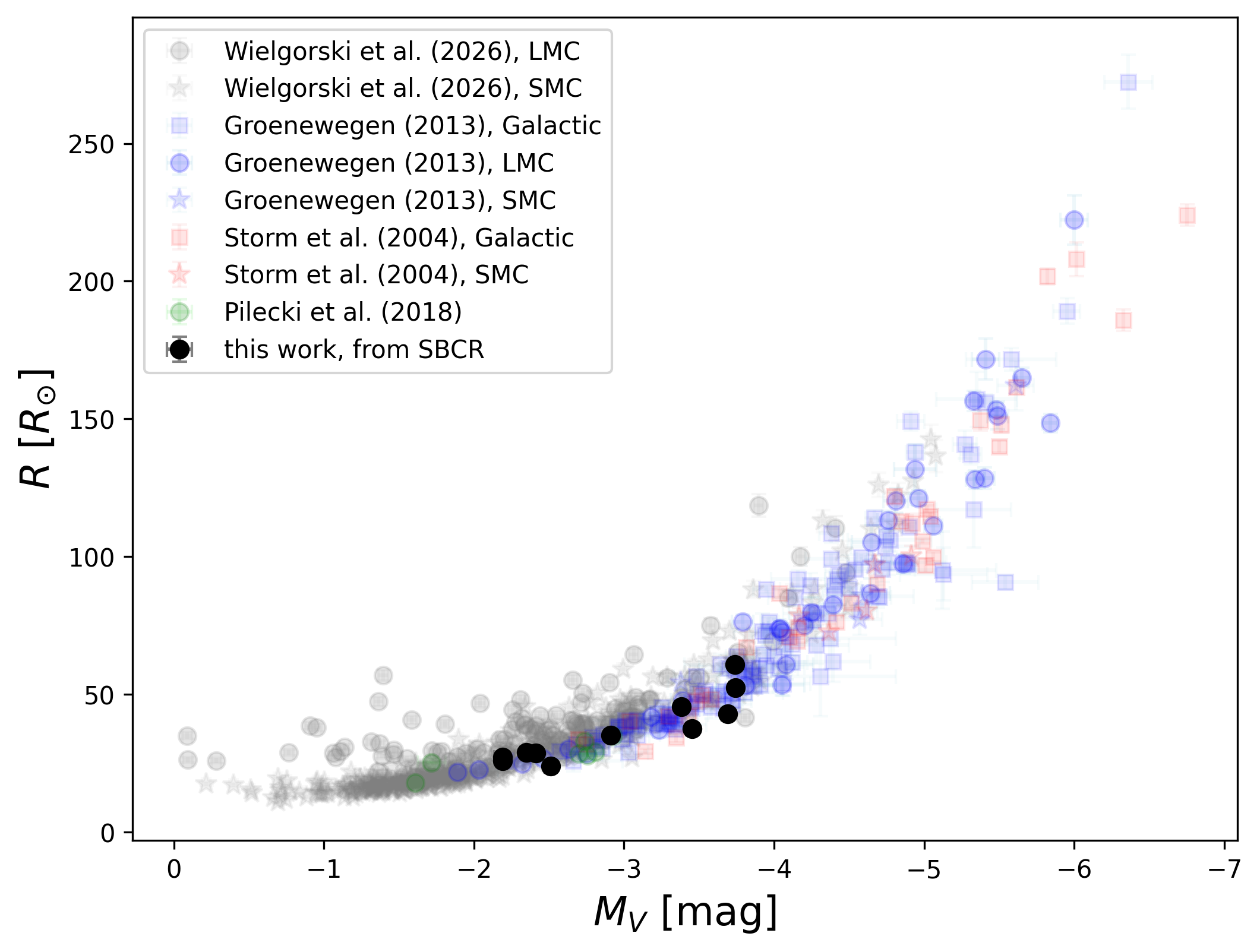}
     \label{fig:r_sub1}
   \end{subfigure}%
   \begin{subfigure}{0.43\textwidth}
    \centering
     \includegraphics[width=\textwidth]{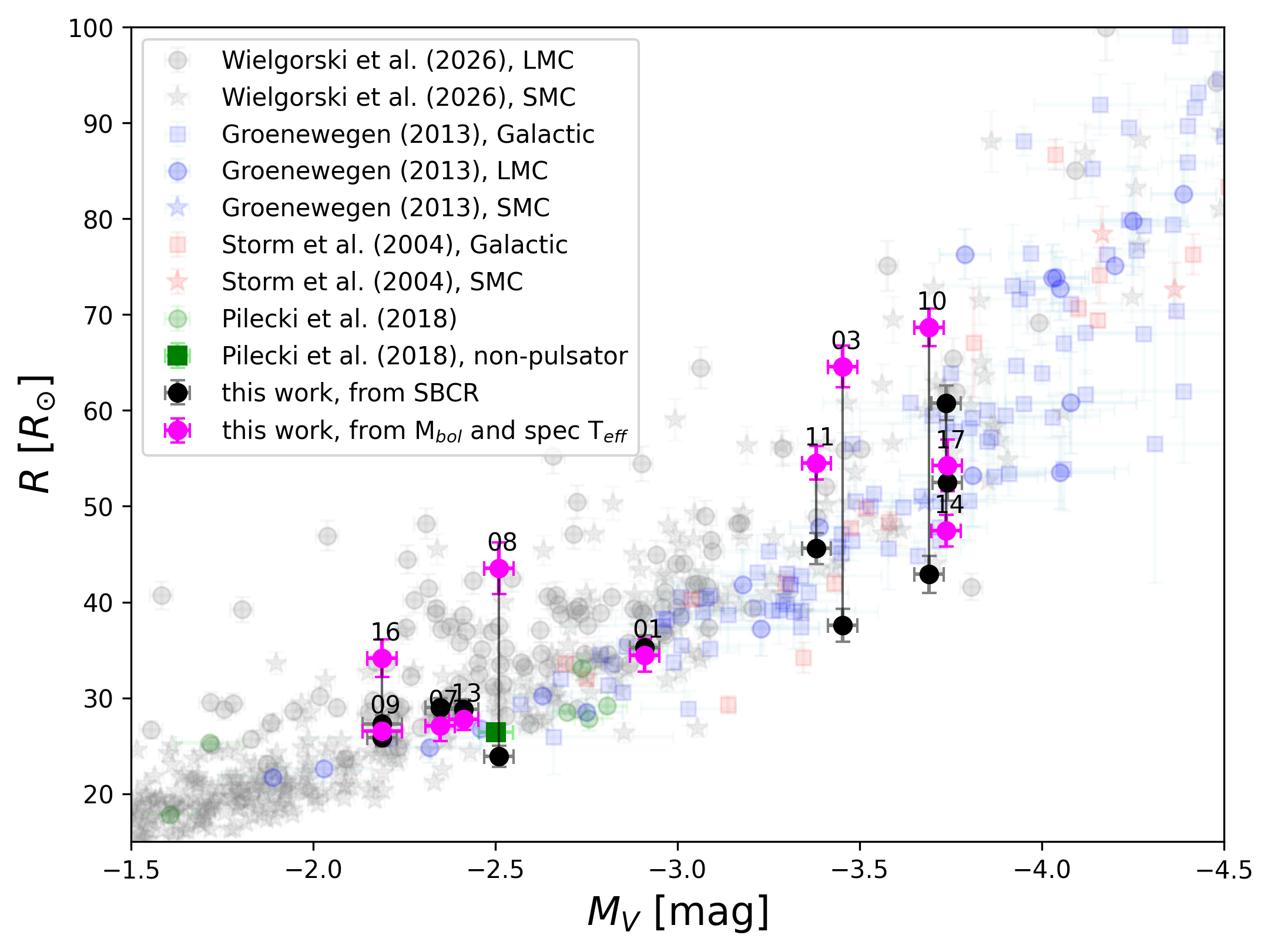}
     \label{fig:r_sub2}
   \end{subfigure}
   \caption{\textit{Left}: Comparison of stellar mean radii of candidates 
             for non-pulsating stars calculated from SBCRs (black points) with 
             Cepheids from the MW (squares), LMC (points), and SMC (star symbols).
             \textit{Right}: Zoom-in of the left panel. Annotated, magenta points
             mark stellar radii calculated directly from absolute bolometric
             magnitudes and $T_{\rm eff}$ from spectroscopy. Corresponding stars 
             are connected with the black lines.
             The non-pulsating star from the binary system from \citet{Pilecki2018}
             residing in the IS is marked with green square.}
   \label{fig:radii_comp}
   \end{figure*}

To compare the radii derived using five independent SBCR calibrations,
in the left panel of Fig.~\ref{fig:radii_comp} we plot the mean radii from
Table~\ref{tab:par_phot} as a~function of the absolute $V$~magnitude (black points), 
together with literature values. 
\citet{Storm2004} provided stellar radii for $34$ Galactic Cepheids and five
metal-poor Cepheids from the Small Magellanic Cloud (SMC), determined using  
the Baade-Wesselink (BW) method (see their Table~3; shown as red points 
and star symbols in Fig.~\ref{fig:radii_comp}). 
\citet{Groenewegen2013} derived radii for a~sample of $128$ Galactic Cepheids, 
as well as $36$ LMC and six SMC Cepheids, also based on the BW method (see 
their Table~10; blue squares, points, and star symbols in Fig.~\ref{fig:radii_comp}). 
Wielgorski et al. (in preparation) calculated the radii using SBCRs for
$138$ classical F Cepheids in the LMC and for $519$ Cepheids 
in the SMC (marked in Fig.~\ref{fig:radii_comp} with gray squares and star symbols).
Finally, LMC Cepheids in binary systems analyzed by \citet{Pilecki2018} are 
shown as green points, while one confirmed non-pulsating companion residing
in the LMC IS is marked with a~green square.
The radii obtained for our candidates using five different SBCRs, as listed 
in Table~\ref{tab:par_phot}, agree well within the errors and follow the general 
trend of literature values for Cepheids.  

For an additional comparison, in the right panel of Fig.~\ref{fig:radii_comp} 
we show stellar radii determined directly from the Stefan--Boltzmann law using 
the absolute bolometric magnitudes, $M_{bol}$, calculated from $M_V$
(as described in the previous section) and spectroscopic $T_{\rm eff}$ (see
Table~\ref{tab:par_phot}). 
The bolometric corrections ($BC_V$) were computed using the calibration
of \citet{Alonso1999} and the spectroscopic values of $T_{\rm eff}$ and [M/H].
In general, there is a~good agreement between these values, with a~few exceptions. 
The radii of CAN-08 and CAN-16 estimated with the second method are clearly 
overestimated. We suspect that these two stars may in fact be binaries 
(see the next section for discussion), which could affect the determination of
the spectroscopic $T_{\rm eff}$, as they were treated as single objects.
Three additional stars with clearly overestimated radii are CAN-03, CAN-10, and
also CAN-11. The first two stars are located close to the blue edge of the IS
(see Fig.~\ref{fig:cmd_comp}).
CAN-10 is characterized by relatively high rotation, whereas CAN-03, in contrast,
has the lowest rotation velocity in the sample. 
CAN-11, on the other hand, is located at the center of the IS and is the second
slowest-rotating object in the sample under study.
The origin of the discrepancy between the radii derived using different methods 
for these stars remains unclear.

%-------------------------------------------------------------
%-------------------------------------------------------------

\subsection{Results from the spectroscopy} 
\label{ssec:res_spec}

\begin{figure*}[ht!]
   \centering
    \includegraphics[width=0.83\textwidth]{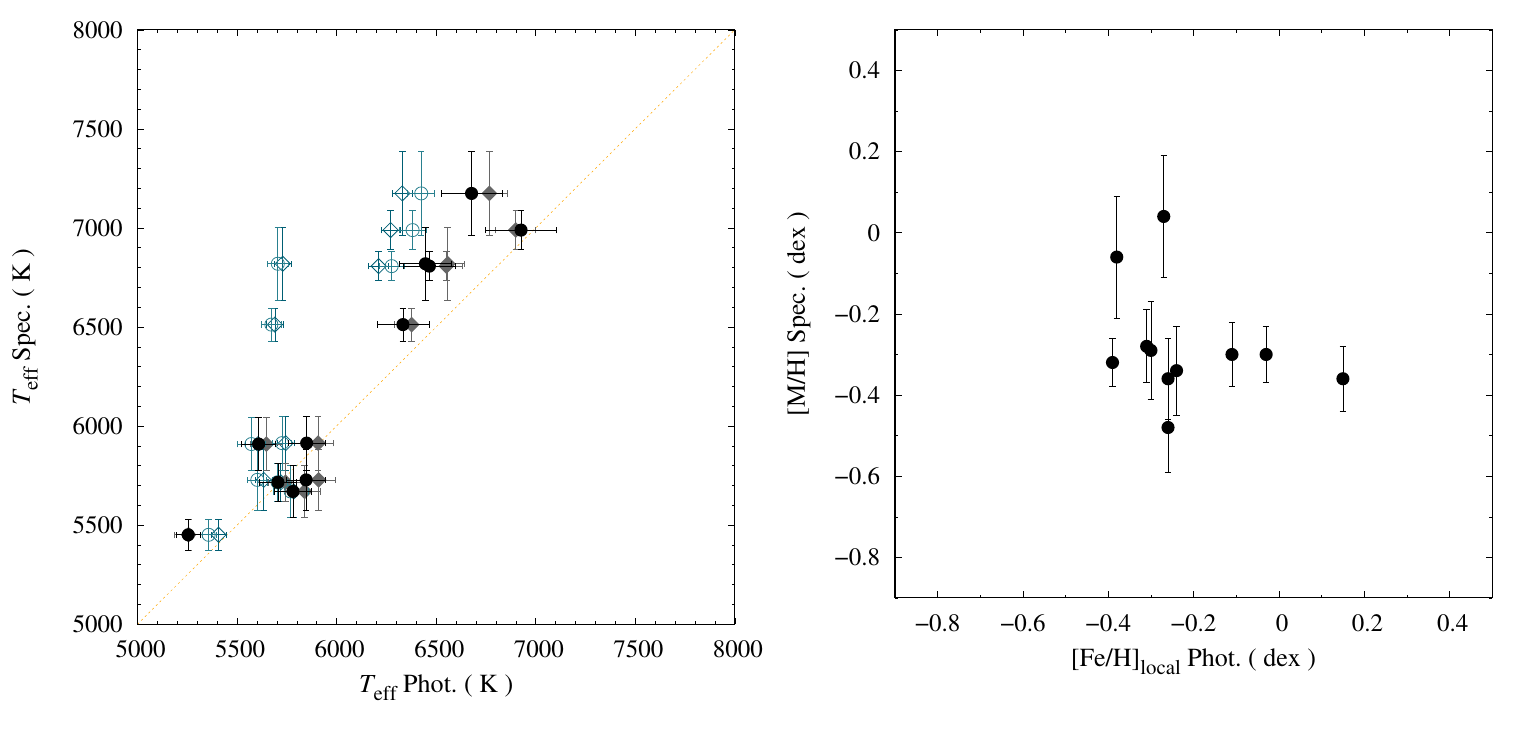}
      \caption{Comparison of $T_{\rm eff}$ (\textit{left}) and [M/H] (\textit{right})
               for all objects, obtained using two methods: analysis of photometry 
               and spectral synthesis with the GSSP code.
               In the left panel, filled black circles represent $T_{\rm eff}$ 
               derived from the calibration of \citet{RamirezMelendez2005} using 
               the $(b-y)$ color, while open turquoise circles show the same for the 
               $(V-K)$ color. Filled gray diamonds denote $T_{\rm eff}$ from 
               calibration of \citet{Alonso1999} for the $(b-y)$ color and open turquoise 
               diamonds show the corresponding values for the $(V-K)$ color.
               In the right panel, photometric metallicities are calculated for 
               the field red giants from the vicinity of our targets, calculated 
               based on the Str\"omgren photometry, while spectroscopic metallicities
               are derived for individual candidates from available spectra.
               This comparison indicates that, in terms of metallicity, the candidates 
               do not differ from field stars.  
               Error bars indicate the $1 \sigma$ uncertainties.
               Dashed orange lines represent 1:1 relations.}
         \label{fig:fig_comp_All-par}
   \end{figure*}

Table~\ref{tab:t_res} presents the results obtained with the spectral synthesis 
using the GSSP code. 
A~comparison of the atmospheric parameters, $T_{\rm eff}$ and ${\rm [Fe/H]}_{local}$,
derived from photometry (see Sects.~\ref{ssec:analysis_phot} and \ref{ssec:t_eff},
summarized in Tables~\ref{tab:par_phot} and \ref{tab:phot_sum}) with $T_{\rm eff}$
and [M/H] derived from spectroscopy is shown in Fig.~\ref{fig:fig_comp_All-par}.
The values agree with the photometric estimates within the uncertainties.

%-------------------------------------------------------------
%-------------------------------------------------------------

In particular, the agreement in temperature values supports the conclusion 
based on optical photometry (presented in Paper~1) that these objects are 
indeed located within the Cepheid IS.
The spectroscopic parameters agree with the previous photometric estimates
within the uncertainties. In particular, the consistency between the spectroscopic
and photometric temperature determinations support the conclusion from Paper~I
that these objects are located within the Cepheid IS. Since one of the aims of
the spectroscopic analysis was to test whether their position within the IS could
result from photometric uncertainties or inaccurate reddening corrections, the
agreement between the two independent methods argues against such systematic effects
and suggests that their observed behavior is linked to an underlying physical
mechanism.
This deepens the puzzle of why these stars do not pulsate, despite their 
parameters being consistent with those observed in classical Cepheids 
in the LMC. 

The sample of classical Cepheids in the LMC cluster NGC~1866, analyzed 
spectroscopically by \citet{Lemasle2017}, shows atmospheric parameters
($T_{\rm eff}$, $\log g$, [M/H], and $\xi_{\rm t}$) that closely match the
values derived for our stars. The metallicities of our entire sample are
subsolar, with a~median [M/H] $\approx$$-0.3$~dex, in agreement with typical
values for LMC Cepheids \citep{Romaniello2022,Hocde2023}.
Furthermore, their agreement with the metallicities derived for red giant
stars from the vicinity of the candidates from Str\"omgren photometry indicates
that the candidates do not differ in metallicity from the surrounding
field-star population.
The $\log g$ values derived here ($\sim1.0-1.9$~dex) are slightly higher 
than those reported by \citet{Romaniello2022}, but remain consistent within 
the uncertainties. 
They are also in good agreement with the high-precision values obtained from 
orbital and geometric analyses of Cepheids in binary systems \citep{Pilecki2018}.

One possibility to consider is that the presence of a~companion could 
influence the observed properties of these stars. In fact, there is evidence 
suggesting a~possible binary nature for at least two objects, CAN-08 and CAN-16. 
This was initially indicated by the broad and strongly asymmetric profiles 
revealed by the cross-correlation and BF methods
(Sect.~\ref{ssec:data_spec}), and is further supported by their unusually
high $\log g$ values ($\sim1$~dex above the typical range in the sample) 
and [M/H] enhanced by $\sim0.3$~dex -- consistent with what would be expected 
if the spectrum of a~binary system composed of two similar stars were analyzed 
as that of a~single star.

\begin{figure}[ht!]
   \centering
    \includegraphics[width=\columnwidth]{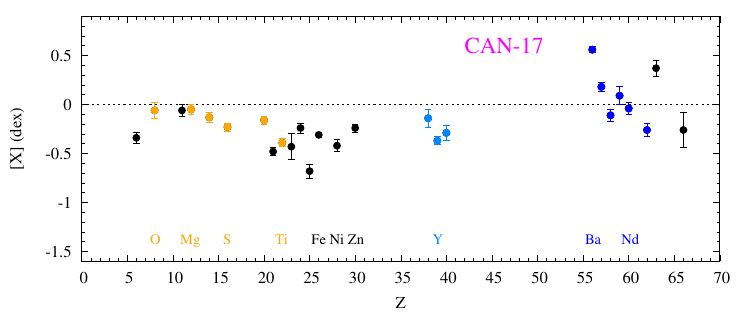}
      \caption{Abundances derived for CAN-17. 
               Elements are shown in black, with $\alpha$ elements highlighted 
               in orange, heavy $s$-process elements in blue, and light $s$-process 
               elements in light blue.}
         \label{fig:fig_X-Z_can-17}
   \end{figure}

Abundances of chemical elements were measured for all objects in the sample
(summarized in Tables~\ref{tab:chem_sum} and \ref{tab:chem_sum2}).
Elemental abundances relative to the Sun, [$X$], based on the solar composition 
of \citet{Asplund2009}, \citet{Scott2015a,Scott2015b}, and \citet{Grevesse2015}, 
are shown for CAN-17 in Fig.~\ref{fig:fig_X-Z_can-17} as an example. 
For all objects, abundances of $\alpha$- and $s$-process elements were determined.

%-------------------------------------------------------------
%-------------------------------------------------------------

The abundances of Fe and $\alpha$ elements are useful for studying the 
chemical evolution of galaxies and the formation of stellar populations,
due to the well-known differences in the lifetimes of the objects that
produce these elements and enrich the interstellar medium. We calculated
[$\alpha$/Fe] using the abundances of Ti, Ca, Si, Mg, and S. The 
[$\alpha$/Fe] values in our sample range from $\sim0.05$ to $0.25$~dex
(see Table~\ref{tab:t_res}).

\begin{figure}[ht!]
   \centering
    \includegraphics[width=\columnwidth]{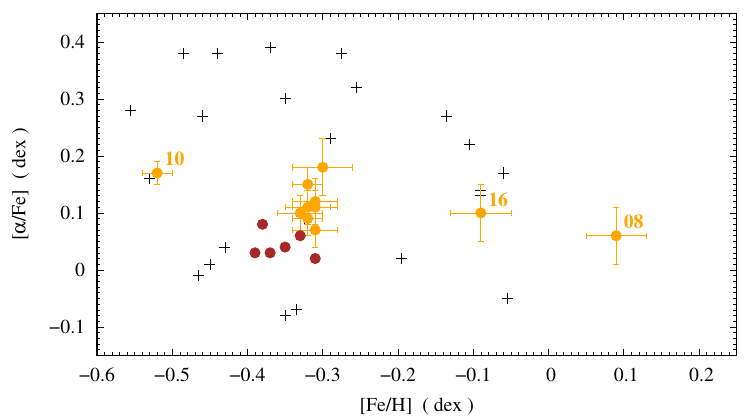}
      \caption{Plot of [$\alpha$/Fe] vs. [Fe/H]. Our sample stars are shown in 
               orange, while classical Cepheids from the LMC cluster NGC~1866 
               \citep{Lemasle2017} are shown in brown. Plus signs indicate Galactic 
               barium stars from \citet{AllenBarbuy2006}.}
         \label{fig:fig_alfFe-FeH}
   \end{figure}

High-resolution spectroscopic studies of classical Cepheids in the LMC 
indicate moderate $\alpha$-element abundances ([$\alpha$/Fe] $\gtrsim 0$), 
consistent with a~young disk population rather than an old, high-$\alpha$ 
population \citep{Lemasle2017,Romaniello2022}. The [$\alpha$/Fe] ratios
in our sample show only a~mildly super-solar trend (Fig.~\ref{fig:fig_alfFe-FeH}), 
in good agreement with [$\alpha$/Fe] observed in barium stars at similar 
[Fe/H] \citep[see Fig.~19 from][]{deCastro2016}.

%-------------------------------------------------------------
%-------------------------------------------------------------

All parameters of our candidates for non-pulsating stars in the Cepheid
IS agree well with those of classical Cepheids, with the exception of the
$s$-process elements. We measured the abundances of all elements conventionally
used to study $s$-process nucleosynthesis due to their dominant contribution
to the process. \citet{LuckBond1991} defined indices convenient for analyzing
$s$-process element abundances -- [hs] for the heavy (barium-peak) elements
and [ls] for the light (zirconium-peak) elements.

\citet{Busso1995} introduced the [hs/ls] ratio as a~tool for assessing 
the efficiency of the $s$ process, defining [hs] and [ls] as two groups 
of $s$-process peaks, and showed that [hs/ls] provides a~measure of neutron 
exposure, i.e., the efficiency of the $^{13}$C-pocket. This indicator is 
now commonly used to study nucleosynthesis in asymptotic giant branch (AGB) 
stars and to classify chemically peculiar stars, including barium stars. 

Several studies of barium stars over the last few decades have established 
an empirical division into two groups: moderate $s$-process enhanced stars, 
with typical $0.2 \lesssim$ [Ba/Fe] $\lesssim 0.5$ and [hs/ls] $\approx +0.0$ 
to $+0.3$, and strong-barium stars, which reach [Ba/Fe] $\gtrsim +0.6$ and 
[hs/ls] $\gtrsim +0.3$~dex. These trends are consistent with high-resolution 
abundance studies \citep[e.g.,][]{AllenBarbuy2006,Smiljanic2007,Merle2016} 
and homogenized large samples \citep{deCastro2016}.

Different authors adopt slightly different sets of elements to calculate
[hs] and [ls]. In our analysis, we used the abundances of four heavy 
$s$-process elements (Ba, La, Ce, and Nd) to calculate [hs/Fe], and three
light $s$-process elements (Zr, Y, Sr) to calculate [ls/Fe]. 
Additionally, we computed indices [hs$_{\rm{L}}$] and [ls$_{\rm{L}}$] using
only the pairs (La, Nd) and (Zr, Y), based on abundances published for
classical Cepheids in the LMC cluster NGC~1866 by \citet{Lemasle2017}, as 
well as from our data, to allow for a direct comparison (see Table~\ref{tab:t_res}).

\begin{figure*}[ht!]
   \centering 
   \sidecaption
    \includegraphics[width=0.5\textwidth]{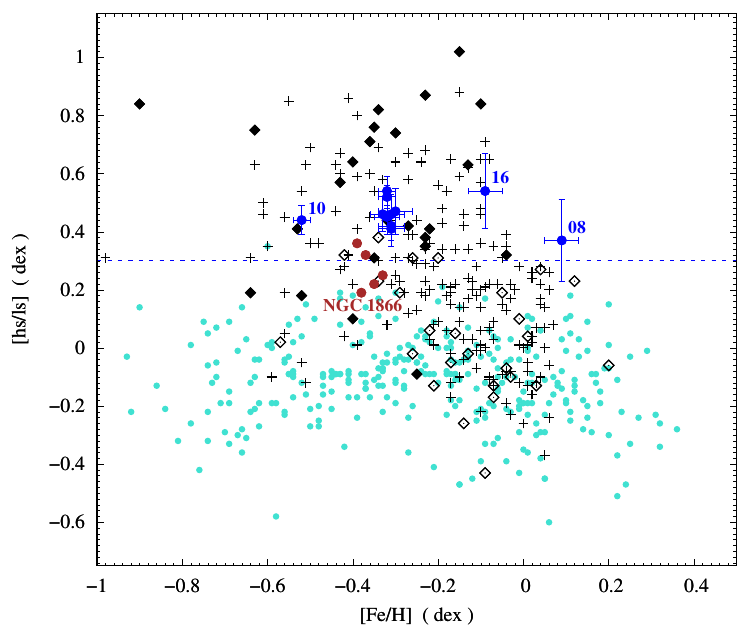}
      \caption{Efficiency of the $s$ process, expressed as [hs/ls], as a~function 
               of metallicity [Fe/H]. Our sample stars are shown in filled blue 
               circles with $1 \sigma$ error bars for [hs$_{\rm L}$/ls$_{\rm L}$].
               Classical Cepheids from the LMC cluster NGC~1866 \citep{Lemasle2017} 
               are shown in brown. Plus signs indicate barium stars from 
               \citet{deCastro2016}, and diamonds indicate barium stars from 
               \citet{Jorissen2019}: filled for strong- and open for mild-barium 
               stars, respectively. 
               Cyan small circles mark Cepheids from \citet{Trentin2026}.
               The horizontal dashed line at $\rm [hs/ls] = +0.3$~dex marks the 
               empirical boundary separating strong- and mild-barium stars.}
         \label{fig:fig_hsls-FeH}
   \end{figure*}

Figure~\ref{fig:fig_hsls-FeH} shows the $s$-process efficiency indicator 
[hs$_{\rm L}$/ls$_{\rm L}$] as a~function of metallicity [Fe/H] for our sample,
compared with the Cepheid sample of \citet{Lemasle2017} and \citet{Trentin2026},
as well as barium stars from \citet{deCastro2016} and \citet{Jorissen2019}.
From the original sample of $340$ stars from \citet{Trentin2026}, we retained 
$314$ Cepheids with reliable measurements of the abundances of interest. 
The stars in our sample lie close to, but above, the lower limit of 
[hs/ls] $> 0.3$~dex and, according to this criterion, could be classified 
as strong-barium stars.

%-------------------------------------------------------------
%-------------------------------------------------------------

\begin{figure*}[ht!]
   \centering 
   \sidecaption
    \includegraphics[width=0.5\textwidth]{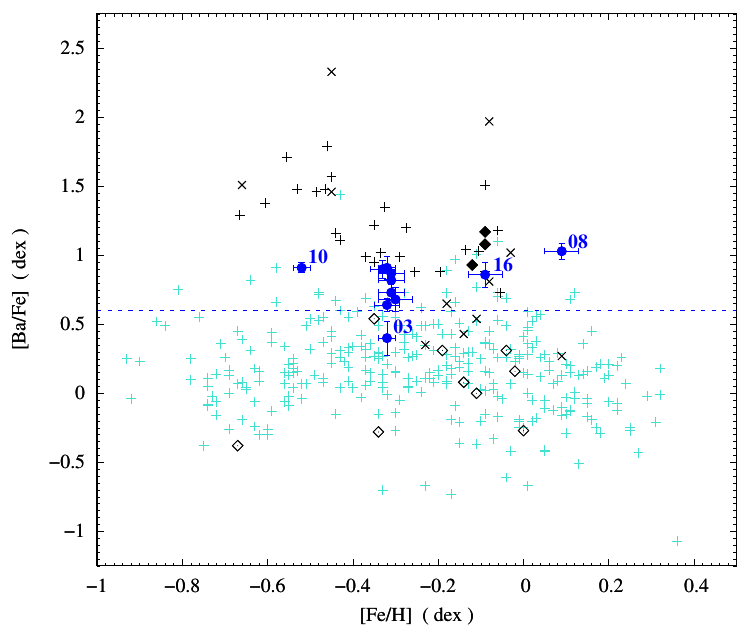}
      \caption{Plot of [Ba/Fe] vs. [Fe/H]. Objects from our sample, shown with blue 
               circles with $1 \sigma$ error bars, are compared to Galactic barium
               stars from \citet{AllenBarbuy2006} (plus signs), \citet{Merle2016} 
               (crosses), and  \citet{Smiljanic2007} (diamonds): filled for strong- 
               and open for mild-barium stars, respectively.
               Cyan crosses mark Cepheids from \citet{Trentin2026}.
               The horizontal dashed line at [Ba/Fe] $= +0.6$~dex marks the empirical 
               boundary separating strong- and mild-barium stars.}
         \label{fig:fig_BaFe-FeH}
   \end{figure*}

Figure~\ref{fig:fig_BaFe-FeH} shows [Ba/Fe] as a~function of metallicity.
All objects in our sample exhibit significant barium enhancement. Based on the 
nominal abundances, $10$ out of $11$ stars can be classified as strong-barium 
stars ([Ba/Fe] $> +0.6$), while CAN-03 falls in the mild-barium regime. 
Considering the $3\sigma$ uncertainties, several objects have error bars that 
overlap the conventional boundary between mild- and strong-barium stars.

%-------------------------------------------------------------
%-------------------------------------------------------------

\citet{Jorissen1998} showed that binarity is the norm for barium stars, with 
a~frequency $\gtrsim 90\%$. Barium stars are giants that contain chemical
elements produced by $s$-process nucleosynthesis, even though they are not
evolved enough to activate the $s$-process in their interiors. The $s$-process 
enrichment in these giants is expected to result from accretion from a~companion 
that was formerly on the AGB and is now a~white dwarf (WD). 

The results of our abundance analysis indicate that the stars in our sample
meet the chemical composition criteria for barium stars. It can be postulated 
that these objects could also be binary systems, with the giant stars having 
their chemical composition modified -- enriched in the products of $s$-process 
nucleosynthesis -- through accretion from a~companion. The stars in our sample 
have somewhat higher effective temperatures than typical barium stars (most 
of which have $T_{\rm{eff}} < 5500$~K); however, they may represent hotter 
counterparts that could be described as warm barium stars. Rare cases of barium 
stars with $T_{\rm{eff}}$ overlapping the temperature range of our sample exist; 
for example, HD~204075 \citep{Smiljanic2007, Jorissen2019}, $\zeta$~Cap, 63~Eri, 
14~Aur \citep{Merle2016}, and about two thirds of the $26$-star sample studied 
by \citet{AllenBarbuy2006}.

Additional spectra would be desirable to address the question of the binary
nature of these objects. \citet{Jorissen2019} conducted long-term spectroscopic
monitoring, initiated in 1984 with the CORAVEL spectrograph and later
continued with the more accurate HERMES spectrograph, and derived
spectroscopic orbits for about three dozen barium stars and S stars without
technetium. They determined that the masses of barium stars range from $1$ to 
$3 M_\odot$, with a~tail extending up to $5 M_\odot$ for mild-barium stars, 
whose WD companions have masses between $0.5$ and $1 M_\odot$. These objects 
show radial velocity variations with amplitudes of typically a~few
kilometers per second.
and orbital periods ranging from several months to several years, with some 
extreme cases extending to decades. A~weak correlation was found, indicating 
that mild-barium stars tend to occur in systems with longer orbital periods 
than strong-barium stars.

The stars in our sample lie near the transition between strong- and mild-barium 
stars, so long orbital periods of the order of years are expected. Long-term 
spectroscopic monitoring would therefore be required to test the binary hypothesis 
for our sample, with a radial velocity precision of at least hundredths of
meters per second to enable a~detailed study of orbital and physical parameters.
A~few objects in our sample have two or three spectra separated by $\sim 1- 2$~years; 
however, these time intervals, combined with the relatively low precision of 
our MIKE radial velocity measurements ($3\sigma \sim 1 - 3$~km\,s$^{-1}$), 
do not allow us to confirm or rule out the presence of radial velocity variations. 

To address the key question of this work -- why these objects do not pulsate
despite exhibiting atmospheric parameters similar to those of classical
Cepheids -- we note that enhanced $s$-process abundances alone cannot account 
for the absence of variability, as heavy elements do not significantly affect 
the $\kappa$ mechanism driving Cepheid pulsations. A~more plausible explanation 
involves evolutionary effects related to past binary interaction and mass 
transfer from a~former AGB companion, which may have altered the internal 
structure of the stars. In this context, rotation may play a~contributory role. 
Classical Cepheids are known to have low surface rotation due to their large 
radii. 
Typical equatorial rotational velocities given by \citep{BersierBurki1996} 
are lower than $V_{\rm rot} \sim 10$~km\,s$^{-1}$, and an average velocity given 
by \citet{Anderson2014} equals $12.3$~km\,s$^{-1}$. 
Our targets exhibit projected rotational velocities in the range of approximately
$6 - 32$~km\,s$^{-1}$, including values higher than typical literature estimates 
for Cepheid rotational velocities.
Notably, the lowest $V_{\rm{rot}} \sin{i}$ value is observed in the star showing
the weakest $s$-process enrichment (CAN-03), consistent with a~mild-barium star
classification. This behaviour supports a~mass-transfer scenario in which angular
momentum is accreted together with $s$-process-enriched material.  

%%%%%%%%%%%%%%%%%%%%%%%%%%%%%%%%%%%%%%%%%%%%%%%%%%%%%%%%%%%%%% 
\section{Summary and conclusions} 
\label{sec:conclusions}

In this work, we analyzed photometric and spectroscopic data for $11$ 
candidates for non-pulsating stars located in the Cepheid IS in the LMC.
For the photometric determination of $T_{\rm eff}$, we used two calibrations 
based on the $(b-y)$ and $(V-K)$ colors. Stellar radii were calculated using 
five different SBCR calibrations for Cepheids, giants, and supergiants (see 
Table~\ref{tab:par_phot}). 
For the spectral analysis, we applied the spectral synthesis method to derive 
atmospheric parameters; namely, $T_{\rm eff}$, metallicity, surface gravity, 
microturbulent velocity, projected rotational velocity (see Table~\ref{tab:t_res}),
and chemical abundances for up to $\sim30$ elements (see Tables~\ref{tab:chem_sum}
and \ref{tab:chem_sum2}).
The obtained atmospheric parameters indicate that these stars are indeed 
giants located within the Cepheid IS. Their position is therefore not the 
result of photometric errors or incorrect reddening corrections, but rather 
some physical mechanism is responsible for the lack of pulsations. 
Their metallicities are in good agreement with the average metallicity 
of Cepheids in the LMC \citep[about $-0.3$~dex; e.g.,][]{Romaniello2022,Hocde2023}.
The derived stellar radii (see Fig.~\ref{fig:radii_comp}) and $T_{\rm eff}$
(see Fig.~\ref{fig:fig_comp_All-par}) agree well with those reported for
Cepheids in the literature.

The spectral line profiles of CAN-08 and CAN-16, from cross-correlation 
and BF methods, are significantly broader compared to those of the rest of 
the sample.
In addition, their spectroscopically derived atmospheric parameters are
noticeably different from those of the other stars.
Their near-solar metallicities and high projected rotational velocities might 
suggest rapid rotation or, alternatively, indicate a~binary nature of these 
two objects, where the components of the systems would be two similar stars.
It should be noted, however, that binarity does not necessarily imply the
scenario in which the two stars lie outside the IS, as individual components may
still reside within the IS boundaries \citep[see, e.g.,][]{Pilecki2018}.
CAN-16 additionally exhibits asymmetric line profiles, which may indicate 
binarity but could also be caused by the presence of non-radial pulsation modes
with a large $l$ number in that star. 
Unfortunately, only single-epoch spectra are available for both stars, and 
additional observations are required to confirm or refute their binary nature. 
The available data are also insufficient to verify the presence of high-order 
radial modes with large $n$ numbers. 

CAN-17 is also characterized by a~relatively high rotational velocity, which 
is unusual for giant stars. This object is the only one in our sample for 
which three spectra were collected. The spectral lines are broad but symmetric, 
and there is no clear indication of binarity, unless a~potential companion 
is a~WD not detectable in optical spectra. Despite the fact that this 
star lies well within the IS, such a~possibility cannot be ruled out. 

A~common feature of all analyzed candidates is an enhancement of barium heavy 
$s$-process elements, which in most cases is typical of strong-barium stars 
(with the exception of CAN-03, which falls into the mild-barium star regime). 
According to the literature \citep[e.g.,][]{Jorissen1998}, barium stars are 
giants enriched in $s$-process elements as a~result of mass transfer from 
a~companion during its post-AGB phase. Our candidates
meet the criteria for barium stars, although their $T_{\rm eff}$ values are 
somewhat higher than those typically observed in such objects. 
This may suggest that their past mass-transfer history modified their chemical 
composition. While enhanced $s$-process abundances alone cannot explain the
absence of pulsations, a~possible explanation (at least for some of our 
candidates) involves evolutionary effects related to past binary interactions, 
which could have altered the internal structure of the stars. 

This work does not provide a~definitive explanation for the nature of
non-pulsating stars located within the LMC Cepheid IS, nor do the obtained 
results yield a~clear answer regarding the origin of their lack of pulsations. 
However, the analysis strongly suggests that multiple mechanisms may be responsible 
for this phenomenon. At the same time, this study presents a~detailed investigation 
of their photometric and, in particular, spectroscopic properties, thereby 
improving our knowledge about the physical parameters and chemical composition 
of these mysterious objects.

\section{Data availability}
\label{data_ava}

The reduced one-dimensional spectra for our candidates (decribed in Table~\ref{tab:obslog})
are only available in electronic form at the CDS via anonymous ftp to
\url{cdsarc.u-strasbg.fr} (130.79.128.5) or via
\url{http://cdsweb.u-strasbg.fr/cgi-bin/qcat?J/A+A/}.

%%%%%%%%%%%%%%%%%%%%%%%%%%%%%%%%%%%%%%%%%%%%%%%%%%%%%%%%%%%%%%
\begin{acknowledgements}  
      We thank anonymous referees for comments that helped to improve 
      original manuscript. 
      The research leading to these results has received funding from 
      the European Research Council (ERC) under the European Union’s 
      Horizon 2020 research and innovation program (grant agreement No. 
      951549 -- UniverScale). We also acknowledge support from the Polish 
      Ministry of Science and Higher Education grant DIR-WSIB.92.2.2024. 
      We gratefully acknowledge financial support for this work from the 
      BASAL Centro de Astrofisica y Tecnolo-gias Afines (CATA) AFB-170002 
      and the Millenium Institute of Astrophysics (MAS) of the Iniciativa 
      Cientifica Milenio del Ministerio de Economia, Fomento y Turismo de 
      Chile, project IC120009. 
      B.P. acknowledges financial support from the Polish National
      Science Center grant SONATA BIS 2020/38/E/ST9/00486. 
      R.S. is supported by the National Science 
      Center, Poland, OPUS project 2024/53/B/ST9/02630.
      W.G. also gratefully acknowledges support from the ANID BASAL project
      ACE210002.
      This work has made use of data from the Las Campanas Observatory 
      (6.5 m Magellan-Clay, MIKE spectrograph). 
      Software used in this work: IRAF \citep{Tody1986,Tody1993}, Astropy 
      \citep{Astropy2013}, NumPy \citep{Numpy1,Numpy2}, SciPy \citep{Virtanen2020}, 
      Matplotlib \citep{Hunter2007}, extinction \citep{extinction2016}, 
      GSSP \citep{Tkachenko2015}. 
\end{acknowledgements}

\bibliographystyle{aa} 
\bibliography{ms_arXiv} 

\begin{appendix}

\onecolumn
\section{Summary of photometric data for the analyzed candidates}
\label{sec:ap_sum}

%-------------------------------------------------------------
\begin{table*}[ht!] 
\small
\caption{\label{tab:phot_sum} Photometric data for $11$ candidates for
non-pulsating stars analyzed in this work.}
\centering
\begin{tabular}{lccccccccccc}
\hline\hline 
CAN
& 01         & 03         & 07         & 08         & 09         & 10        & 11          & 13                & 14                  & 16         & 17        \\
\hline
$V_S$                                                                   
& 15.958     & 15.407     & 16.518     & 16.290     & 16.560     & 15.100    & 15.469      & 16.463    & 15.082    & 16.557     & 15.035    \\
& $\pm$0.004 & $\pm$0.003 & $\pm$0.005 & $\pm$0.005 & $\pm$0.005 & $\pm$0.003 & $\pm$0.003 & $\pm$0.004 & $\pm$0.03 & $\pm$0.004 & $\pm$0.002 \\ 
$V_{S,2021}$                                                                   
& 15.878     & 15.331     & 16.445     & 16.219     & 16.484     & 15.024    & 15.397      & 16.376    & 14.998    & 16.471     & 14.956    \\
& $\pm$0.004 & $\pm$0.003 & $\pm$0.005 & $\pm$0.005 & $\pm$0.005 & $\pm$0.003 & $\pm$0.003 & $\pm$0.004 & $\pm$0.03 & $\pm$0.004 & $\pm$0.002 \\ 
$V_{OGLE-II}$                                                       
& 15.892     & 15.340     & 16.411     & 16.207     & 16.472     & 15.040     & 15.400     & 16.378    & 15.014     & 16.500     & 15.002     \\ 
& $\pm$0.019 & $\pm$0.012 & $\pm$0.017 & $\pm$0.034 & $\pm$0.026 & $\pm$0.008 & $\pm$0.015 & $\pm$0.019 & $\pm$0.011 & $\pm$0.017 & $\pm$0.010 \\ 
$V_{OGLE-III}$                                                          
& 15.889     & 15.355     & 16.427     & 16.224     & 18.425     & 15.028     & 15.391     & 16.373    & 15.009     & 16.500     & 14.974     \\ 
& $\pm$0.006 & $\pm$0.004 & $\pm$0.007 & $\pm$0.005 & $\pm$2.102 & $\pm$0.006 & $\pm$0.005 & $\pm$0.005 & $\pm$0.005 & $\pm$0.009 & $\pm$0.004 \\ 
$V_{OGLE-III,S}$                                                          
& 15.816     & 15.317     & 16.426     & 16.001     & 16.519     & 14.999     & 15.370     & 16.295    & 15.009     & 16.407     & 15.026     \\ 
& $\pm$0.028 & $\pm$0.020 & $\pm$0.028 & $\pm$0.029 & $\pm$0.036 & $\pm$0.014 & $\pm$0.019 & $\pm$0.027 & $\pm$0.005 & $\pm$0.042 & $\pm$0.016 \\ 
\hline 
$I_{OGLE-II}$                                                        
& 15.103     & 14.721     & 15.654     & 15.709     & 15.699     & 14.456     & 14.655     & 15.594    & 14.141     & 15.884     & 14.225     \\ 
& $\pm$0.008 & $\pm$0.012 & $\pm$0.010 & $\pm$0.022 & $\pm$0.016 & $\pm$0.009 & $\pm$0.016 & $\pm$0.014 & $\pm$0.009 & $\pm$0.017 & $\pm$0.010 \\ 
$I_{OGLE-III}$                                                       
& 15.090     & 14.747     & 15.628     & 15.676     & 15.808     & 14.427     & 14.639     & 15.549    & 14.102     & 15.872     & 14.202     \\ 
& $\pm$0.008 & $\pm$0.006 & $\pm$0.007 & $\pm$0.008 & $\pm$0.008 & $\pm$0.006 & $\pm$0.006 & $\pm$0.007 & $\pm$0.005 & $\pm$0.008 & $\pm$0.007 \\ 
\hline
$K_{s,IRSF}$                                                                    
& 14.130     & 13.970     & 14.600     & 14.960     & 14.750     & 13.640     & 13.600     & 14.610    & 13.020     & 15.080     & 13.027    \\
& $\pm$0.020 & $\pm$0.020 & $\pm$0.020 & $\pm$0.020 & $\pm$0.030 & $\pm$0.010 & $\pm$0.020 & $\pm$0.010 & $\pm$0.010 & $\pm$0.040 & $\pm$0.020 \\
$K_{VMC}$                                                                    
& 14.125     & 13.895     & 14.575     & 14.860     & 14.708     & 13.608     & 13.581     & 14.573    & 13.013     & 14.792     & 13.255    \\
& $\pm$0.004 & $\pm$0.004 & $\pm$0.006 & $\pm$0.007 & $\pm$0.006 & $\pm$0.003 & $\pm$0.003 & $\pm$0.006 & $\pm$0.003 & $\pm$0.007 & $\pm$0.003 \\
\hline 
$(b-y)$                                                                    
& 0.439      & 0.286      & 0.421      & 0.274      & 0.456      & 0.302      & 0.350     & 0.462      & 0.542      & 0.291      & 0.409    \\ 
& $\pm$0.006 & $\pm$0.004 & $\pm$0.007 & $\pm$0.005 & $\pm$0.007 & $\pm$0.004 & $\pm$0.004 & $\pm$0.006 & $\pm$0.004 & $\pm$0.006 & $\pm$0.003 \\ 
$m1$                                                                    
& 0.231      & 0.155      & 0.158      & 0.173      & 0.239      & 0.135      & 0.250     & 0.270      & 0.379      & 0.195      & 0.315    \\ 
& $\pm$0.009 & $\pm$0.007 & $\pm$0.010 & $\pm$0.010 & $\pm$0.011 & $\pm$0.005 & $\pm$0.007 & $\pm$0.010 & $\pm$0.008 & $\pm$0.010 & $\pm$0.005 \\
$c1$                                                                    
& 0.653      & 1.144     & 0.486      & 0.955      & 0.567      & 1.097      & 0.941     & 0.577      & 0.533      & 0.881      & 0.689    \\ 
& $\pm$0.008 & $\pm$0.007 & $\pm$0.010 & $\pm$0.010 & $\pm$0.012 & $\pm$0.005 & $\pm$0.008 & $\pm$0.011 & $\pm$0.009 & $\pm$0.010 & $\pm$0.005 \\
\hline
$E(V-I)_{S21}$                                                                    
& 0.128      & 0.137      & 0.137      & 0.081      & 0.080      & 0.072      & 0.124      & 0.124     & 0.100      & 0.060      & 0.076     \\ 
$E(B-V)_{G20}$                                                           
& 0.136      & 0.135      & 0.135      & 0.117      & 0.092      & 0.100      & 0.123      & 0.104     & 0.110      & 0.103      & 0.105     \\ 
$E(B-V)_{G20}'$                                                       
& 0.140      & 0.141      & 0.111      & 0.114      & 0.093      & 0.104      & 0.120      & 0.131     & 0.114      & 0.092      & 0.100     \\ 
$E(B-V)_{N26}$ 
& 0.062      & 0.063      & 0.063      & 0.064      & 0.063      & 0.065      & 0.063      & 0.064     & 0.063      & 0.060      & 0.064     \\ 
$E(B-V)_{ave}$                                                
& 0.100      & 0.103      & 0.093      & 0.080      & 0.072      & 0.075      & 0.092      & 0.096     & 0.084      & 0.066      & 0.074     \\ 
\hline
$\mathrm{[Fe/H]}_{field}$                                                          
& -0.13      & -0.13      & -0.12      & -0.12      & -0.26      & -0.12      & -0.26      & -0.26      & -0.26      & -0.23      & -0.23 \\
$\mathrm{[Fe/H]}_{local}$                                                   
& -0.31      & 0.15       & -0.30      & -0.27      & -0.26      & -0.26      & -0.03      & -0.11      & -0.39      & -0.38      & -0.24 \\
\hline
\end{tabular}
\tablefoot{ \\ 
$V_S$ -- $V$~magnitude from Table~3 of \citet{Narloch2019} calibrated 
from the Str\"omgren photometry. \\ 
$V_{S,2021}$ -- $V$~magnitude from the Str\"omgren photometry calibrated 
using calibration equation from \citet{Narloch2021}. \\ 
$V_{OGLEII}$, $I_{OGLEII}$ -- OGLE-II magnitudes from \citet{Udalski2000}. \\
$V_{OGLEIII}$, $I_{OGLEIII}$ --  OGLE-III magnitudes from \citet{Udalski2008}. \\
$V_{OGLEIII,S}$ -- $V$~magnitude from LMC photometric maps from OGLE-III shallow survey 
\citep{Ulaczyk2012}. \\ 
$K_{s,IRSF}$ -- $K_s$~magnitude from the IRSF Magellanic Clouds Point Source Catalog 
\citep{Kato2007}. \\
$K_{VMC}$ -- $K$~magnitude from the VISTA survey of the Magellanic Clouds system  
\citep[VMC,][]{Cioni2011} for aperture~3. \\ 
$(b-y)$, $m1$, $c1$ -- Str\"omgren indices from Paper~I (see their Table~3). \\ 
$E(V-I)_{S21}$ -- reddening value of $E(V-I)_{median}$ from \citet{Skowron2021}.  \\
$E(B-V)_{G20}$ -- reddening value $E(B-V)$ of the field for $3$~arcmin resolution from 
\citet{Gorski2020}. \\
$E(B-V)_{G20}'$ -- reddening value $E(B-V)$ calculated specifically for each star 
for $3$~arcmin resolution from \citet{Gorski2020}. \\ 
$E(B-V)_{N26}$ -- reddening value $E(B-V)$ calculated specifically for each star 
for $3$~arcmin resolution from \citet{Netzel2026}. \\ 
$E(B-V)_{ave}$ -- averaged reddening calculated as 
$(E(B-V)_{G20}' + E(V-I)_{S21}/1.318 + E(B-V)_{N26})/3$. \\ 
$\mathrm{[Fe/H]}_{field}$ -- metallicity from the Str\"omgren $m1$ index based on 
calibration from \citet{Hilker2000}, averaged for stars from the whole field containing 
a~given candidate. \\
$\mathrm{[Fe/H]}_{local}$ -- metallicity as above, averaged for stars from
the radius of about $50$~arcsec ($350$~pixels for SOI camera) around a~given candidate. \\

}
\end{table*}

\FloatBarrier 
\twocolumn

\onecolumn
\section{Comparison of multiple spectra collected for four candidates for non-pulsating 
stars}
\label{sec:multi_spec}

%-------------------------------------------------------------
\begin{figure}[ht!]
\centering

\begin{subfigure}{0.9\linewidth}
    \centering
    \includegraphics[width=\linewidth]{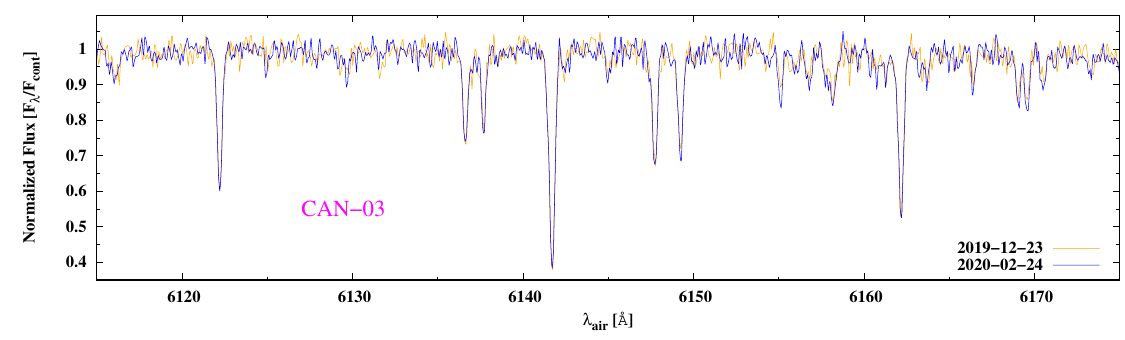}
\end{subfigure}

\vspace{2mm}

\begin{subfigure}{0.9\linewidth}
    \centering
    \includegraphics[width=\linewidth]{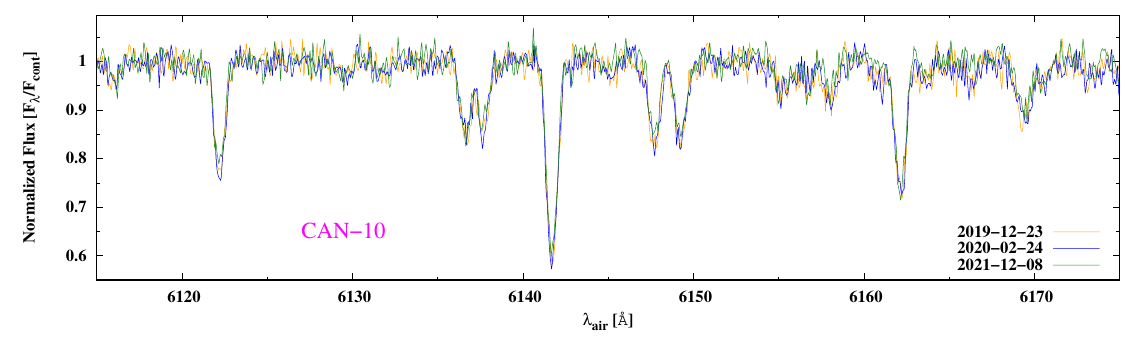}
\end{subfigure}

\vspace{2mm}

\begin{subfigure}{0.9\linewidth}
    \centering
    \includegraphics[width=\linewidth]{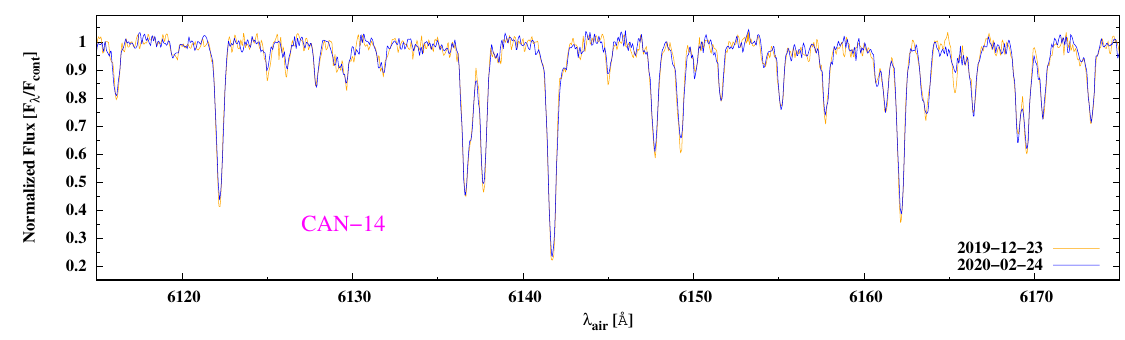}
\end{subfigure}

\vspace{2mm}

\begin{subfigure}{0.9\linewidth}
    \centering
    \includegraphics[width=\linewidth]{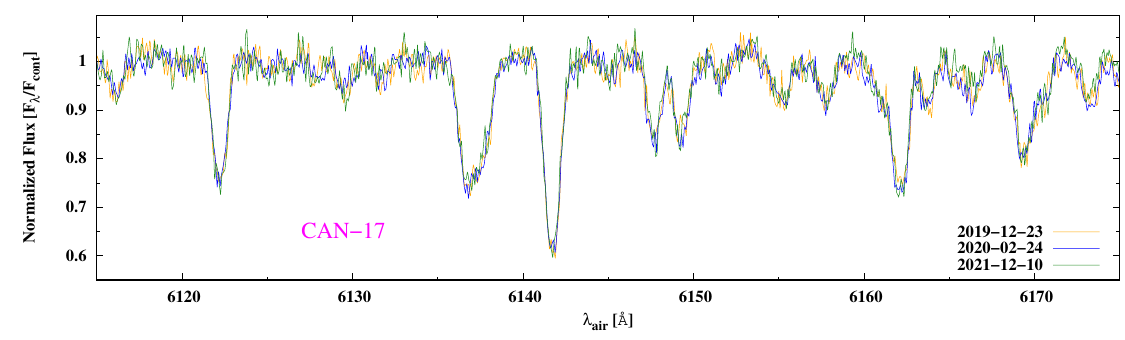}
\end{subfigure}

\caption{Comparison of individual spectra collected for CAN-03, CAN-10,
CAN-14, and CAN-17 in the region around of Ba~II $\lambda 6141.7$~\AA.}
\label{fig:multi_spec}

\end{figure}
%-------------------------------------------------------------

\FloatBarrier 
\twocolumn

\onecolumn 
\section{Observing log}
\label{sec:obslog}

%-------------------------------------------------------------
\begin{table}[ht!] 
\small
\caption{\label{tab:obslog} Log of MIKE spectroscopic observations.}
\centering
\begin{tabular}{lcccc}
\hline\hline
Name & No. of obs. & Date & Radial Vel. & S/N \\
 & & & (km/s) & \\
\hline
CAN-01 & 1 & 2019-12-23 & 267.92 $\pm$ 0.30 & 45 \\ 
CAN-03 & 2 & 2019-12-23 & 271.47 $\pm$ 0.38 & 70 \\
       &   & 2020-02-24 & 271.36 $\pm$ 0.43 &    \\
CAN-07 & 1 & 2019-12-24 & 241.66 $\pm$ 0.26 & 40 \\ 
CAN-08 & 1 & 2019-12-23 & 252.96 $\pm$ 1.26 & 43 \\ 
CAN-09 & 1 & 2019-12-24 & 268.21 $\pm$ 0.34 & 45 \\ 
CAN-10 & 3 & 2019-12-23 & 285.74 $\pm$ 0.97 & 95 \\ 
       &   & 2020-02-24 & 285.28 $\pm$ 0.99 &    \\ 
       &   & 2021-12-08 & 284.81 $\pm$ 1.09 &    \\ 
CAN-11 & 1 & 2019-12-23 & 244.08 $\pm$ 0.29 & 48 \\ 
CAN-13 & 1 & 2019-12-24 & 265.43 $\pm$ 0.34 & 43 \\ 
CAN-14 & 2 & 2019-12-23 & 259.37 $\pm$ 0.38 & 65 \\ 
       &   & 2020-02-24 & 259.70 $\pm$ 0.40 &    \\ 
CAN-16 & 1 & 2019-12-24 & 278.71 $\pm$ 0.78 & 40 \\ 
CAN-17 & 3 & 2019-12-23 & 263.62 $\pm$ 0.60 & 90 \\ 
       &   & 2020-02-24 & 263.33 $\pm$ 0.60 &    \\ 
       &   & 2021-12-10 & 263.12 $\pm$ 0.48 &    \\ 
\hline
\end{tabular}
\end{table}

\section{Results from the equivalent-width method}
\label{sec:ew}

We derived atmospheric parameters for eight candidates from our sample
using the EW method. The resulting parameters are consistent, within the 
uncertainties, with those obtained from the spectral synthesis method.
However, the associated uncertainties are relatively large compared to other 
method, which limits the significance of the constraints derived from them. 
We therefore consider the EW method unsuitable for our purposes and do not 
use its results in the analysis.

\begin{table}[ht!]
\fontsize{9pt}{9pt}\selectfont
\begin{center}
\caption{\small{Parameters derived using the EW method.}}
\label{tab:t_ew} 
\begin{tabular}{@{}l@{\hskip 2mm}l@{\hskip 2mm}l@{\hskip 2mm}l@{\hskip 2mm}l@{\hskip 2mm}l@{}}
\hline
\hline
&&&&&\\
Star    &$T_{\rm{eff}}$ &$\log{g}$      &$\xi_{\rm t}$  &[M/H]          &[$\alpha$/Fe]\\
&&&&&\\
                &[km/s]         &[dex]          &[km/s]         &[dex]          &[dex]        \\
\hline
&&&&&\\
CAN-01          &6030$\pm$330   &2.23$\pm$0.69  &4.39$\pm$0.57  &-0.08$\pm$0.23 &-0.04        \\
&&&&&\\
CAN-03          &7220$\pm$600   &2.08$\pm$0.75  &2.63$\pm$0.39  &0.04$\pm$0.21  &+0.02        \\
&&&&&\\
CAN-07          &6040$\pm$455   &2.55$\pm$0.97  &3.99$\pm$0.70  &-0.17$\pm$0.25 &-0.08        \\
&&&&&\\
CAN-09          &6080$\pm$335   &2.78$\pm$0.70  &4.06$\pm$0.47  &-0.05$\pm$0.25 &-0.03        \\
&&&&&\\
CAN-10          &6845$\pm$405   &1.90$\pm$0.60  &4.10$\pm$0.62  &-0.33$\pm$0.16 &+0.08        \\
&&&&&\\
CAN-11          &6410$\pm$375   &1.55$\pm$0.62  &3.33$\pm$0.40  &-0.14$\pm$0.17 &+0.04        \\
&&&&&\\
CAN-13          &6115$\pm$380   &2.55$\pm$0.76  &5.66$\pm$1.75  &-0.14$\pm$0.23 &-0.16        \\
&&&&&\\
CAN-14          &5770$\pm$185   &1.64$\pm$0.43  &3.92$\pm$0.36  &+0.01$\pm$0.17 &-0.07        \\
\hline
\hline
\end{tabular}
\end{center}
\end{table}

\FloatBarrier 
\twocolumn

\onecolumn
\begin{landscape}

\section{Tables with abundances of elements for the candidates}
\label{sec:chem_sum}

\small
\begin{longtable}{llcccccccccccccccccc} 
\caption{Abundances of elements for candidates for non-pulsating stars.} 
\label{tab:chem_sum} \\
\hline\hline
 &  & CAN-01 & & & CAN-03 & & & CAN-07 & & & CAN-08 & & & CAN-09 & & & CAN-10 & & \\
Z & Element & $\rm log \epsilon (X)$ & $\pm 1\sigma$ & [$X$] & $\rm log \epsilon (X)$ & $\pm 1\sigma$ & [$X$] & $\rm log \epsilon (X)$ & $\pm 1\sigma$ & [$X$] & $\rm log \epsilon (X)$ & $\pm 1\sigma$ & [$X$] & $\rm log \epsilon (X)$ & $\pm 1\sigma$ & [$X$] & $\rm log \epsilon (X)$ & $\pm 1\sigma$ & [$X$] \\
\hline
\endfirsthead
\caption{continued.}\\
\hline\hline
 &  & CAN-01 & & & CAN-03 & & & CAN-07 & & & CAN-08 & & & CAN-09 & & & CAN-10 & & \\
Z & Element & $\rm log \epsilon (X)$ & $\pm 1\sigma$ & [$X$] & $\rm log \epsilon (X)$ & $\pm 1\sigma$ & [$X$] & $\rm log \epsilon (X)$ & $\pm 1\sigma$ & [$X$] & $\rm log \epsilon (X)$ & $\pm 1\sigma$ & [$X$] & $\rm log \epsilon (X)$ & $\pm 1\sigma$ & [$X$] & $\rm log \epsilon (X)$ & $\pm 1\sigma$ & [$X$] \\
\hline
\endhead
\hline
\endfoot
6  & C  & 7.87  & 0.03  & -0.56 & 7.92   & 0.02  & -0.51 & 7.77  & 0.08  & -0.66 & 8.36 & 0.09 & -0.07 & 7.78 & 0.06 & -0.65 & 7.97 & 0.02 & -0.46 \\
7  & N  & -- & -- & -- & 8.18 & 0.04 & 0.35 & -- & -- & -- & 8.41 & 0.22 & 0.58 & -- & -- & -- & 7.51 & 0.23 & -0.32 \\
8  & O  & 8.56 & 0.06 & -0.13 & 8.54 & 0.05 & -0.15 & -- & -- & -- & 8.78 & 0.11 & 0.09 & -- & -- & -- & 8.55 & 0.06 & -0.14 \\
11 & Na & 6.01 & 0.04 & -0.20 & 6.11 & 0.02 & -0.10 & 5.88 & 0.06 & -0.33 & 6.36 & 0.08 & 0.15 & 5.98 & 0.04 & -0.23 & 5.79 & 0.04 & -0.42 \\
12 & Mg & 7.45 & 0.04 & -0.14 & 7.45 & 0.03 & -0.14 & 7.45 & 0.12 & -0.14 & 7.85 & 0.06 & 0.26 & 7.40 & 0.06 & -0.19 & 7.29 & 0.03 & -0.30 \\
13 & Al & -- & -- & -- & -- & -- & -- & -- & -- & -- & -- & -- & -- & 6.38 & 0.04 & -0.05 & -- & -- & -- \\
14 & Si & 7.33 & 0.04 & -0.18 & 7.47 & 0.03 & -0.04 & 7.27 & 0.06 & -0.24 & 7.71 & 0.08 & 0.20 & 7.33 & 0.04 & -0.18 & 7.32 & 0.04 & -0.19 \\
16 & S  & 6.85 & 0.05 & -0.27 & 7.03 & 0.03 & -0.09 & 7.30 & 0.06 & 0.18 & 7.30 & 0.05 & 0.18 & 6.75 & 0.04 & -0.37 & 6.91 & 0.03 & -0.21 \\
20 & Ca & 6.17 & 0.03 & -0.15 & 6.12 & 0.06 & -0.20 & 6.19 & 0.05 & -0.13 & 6.45 & 0.10 & 0.13 & 6.20 & 0.03 & -0.12 & 5.93 & 0.02 & -0.39 \\
21 & Sc & 2.52 & 0.03 & -0.64 & 2.68 & 0.02 & -0.48 & 2.63 & 0.04 & -0.53 & 3.05 & 0.05 & -0.11 & 2.60 & 0.03 & -0.56 & 2.55 & 0.02 & -0.61 \\
22 & Ti & 4.53 & 0.03 & -0.40 & 4.61 & 0.02 & -0.32 & 4.52 & 0.05 & -0.41 & 4.89 & 0.06 & -0.04 & 4.57 & 0.03 & -0.36 & 4.42 & 0.02 & -0.51 \\
23 & V  & 3.56 & 0.06 & -0.33 & 3.73 & 0.04 & -0.16 & 3.37 & 0.12 & -0.52 & -- & -- & -- & 3.44 & 0.06 & -0.45 & 3.36 & 0.08 & -0.53 \\
24 & Cr & 5.36 & 0.03 & -0.26 & 5.43 & 0.03 & -0.19 & 5.30 & 0.05 & -0.32 & 5.85 & 0.06 & 0.23 & 5.36 & 0.04 & -0.26 & 5.24 & 0.03 & -0.38 \\
25 & Mn & 4.78 & 0.04 & -0.65 & 4.93 & 0.03 & -0.49 & 4.77 & 0.05 & -0.65 & 4.88 & 0.14 & -0.54 & 4.74 & 0.04 & -0.68 & 4.72 & 0.04 & -0.70 \\
26 & Fe & 7.15 & 0.03 & -0.32 & 7.15 & 0.02 & -0.32 & 7.17 & 0.04 & -0.30 & 7.56 & 0.04 & 0.09 & 7.14 & 0.03 & -0.33 & 6.95 & 0.02 & -0.52 \\
27 & Co & 4.43 & 0.07 & -0.50 & -- & -- & -- & -- & -- & -- & -- & -- & -- & 4.26 & 0.10 & -0.67 & -- & -- & -- \\
28 & Ni & 5.75 & 0.03 & -0.45 & 5.72 & 0.07 & -0.48 & 5.80 & 0.05 & -0.40 & 5.99 & 0.09 & -0.21 & 5.72 & 0.04 & -0.48 & 5.55 & 0.04 & -0.65 \\
29 & Cu & 3.56 & 0.03 & -0.62 & -- & -- & -- & 3.54 & 0.10 & -0.64 & 4.39 & 0.11 & 0.21 & 3.41 & 0.13 & -0.77 & 3.44 & 0.19 & -0.74 \\
30 & Zn & 3.98 & 0.02 & -0.58 & 4.04 & 0.06 & -0.52 & 3.83 & 0.06 & -0.73 & -- & -- & -- & 3.87 & 0.02 & -0.69 & -- & -- & -- \\
38 & Sr & 2.39 & 0.06 & -0.44 & -- & -- & -- & 2.53 & 0.10 & -0.30 & -- & -- & -- & 2.47 & 0.05 & -0.36 & 2.78 & 0.09 & -0.05 \\
39 & Y  & 1.69 & 0.03 & -0.53 & 1.74 & 0.03 & -0.47 & 1.81 & 0.04 & -0.40 & 2.41 & 0.08 & 0.20 & 1.79 & 0.03 & -0.42 & 1.65 & 0.03 & -0.56 \\
40 & Zr & 2.27 & 0.05 & -0.32 & 2.33 & 0.05 & -0.26 & 1.94 & 0.13 & -0.65 & 2.43 & 0.15 & -0.16 & 2.20 & 0.07 & -0.39 & 2.17 & 0.05 & -0.42 \\
56 & Ba & 2.57 & 0.04 & 0.32 & 2.33 & 0.12 & 0.08 & 2.63 & 0.08 & 0.38 & 3.37 & 0.04 & 1.12 & 2.82 & 0.06 & 0.57 & 2.64 & 0.03 & 0.39 \\
57 & La & 1.14 & 0.03 & 0.03 & 1.30 & 0.03 & 0.19 & 1.20 & 0.05 & 0.09 & 1.41 & 0.15 & 0.30 & 1.20 & 0.04 & 0.09 & 1.06 & 0.04 & -0.05 \\
58 & Ce & 1.45 & 0.04 & -0.13 & 1.57 & 0.04 & -0.01 & 1.45 & 0.07 & -0.14 & 2.10 & 0.11 & 0.52 & 1.50 & 0.04 & -0.08 & 1.40 & 0.05 & -0.18 \\
59 & Pr & 0.53 & 0.06 & -0.19 & 0.64 & 0.10 & -0.08 & 0.80 & 0.09 & 0.08 & -- & -- & -- & 0.69 & 0.07 & -0.03 & -- & -- & -- \\
60 & Nd & 1.31 & 0.04 & -0.11 & 1.41 & 0.04 & -0.01 & 1.42 & 0.06 & 0.00 & 2.05 & 0.13 & 0.63 & 1.37 & 0.05 & -0.05 & 1.20 & 0.08 & -0.22 \\
62 & Sm & 0.72 & 0.03 & -0.23 & 0.78 & 0.08 & -0.17 & 0.71 & 0.08 & -0.25 & -- & -- & -- & 0.75 & 0.04 & -0.20 & 0.61 & 0.14 & -0.34 \\
63 & Eu & 0.92 & 0.03 & 0.40 & 0.28 & 0.10 & -0.24 & 0.90 & 0.07 & 0.38 & -- & -- & -- & 0.92 & 0.03 & 0.40 & -- & -- & -- \\
64 & Gd & 0.66 & 0.15 & -0.42 & -- & -- & -- & -- & -- & -- & -- & -- & -- & -- & -- & -- & -- & -- & -- \\
65 & Tb & 0.08 & 0.20 & -0.23 & -- & -- & -- & -- & -- & -- & -- & -- & -- & -- & -- & -- & -- & -- & -- \\
66 & Dy & 0.90 & 0.10 & -0.20 & -- & -- & -- & -- & -- & -- & -- & -- & -- & 0.94 & 0.12 & -0.16 & 0.70 & 0.30 & -0.40 \\
68 & Er & 0.93 & 0.17 & 0.00 & -- & -- & -- & -- & -- & -- & -- & -- & -- & -- & -- & -- & -- & -- & -- \\
71 & Lu & 0.25 & 0.04 & 0.15 & -- & -- & -- & 0.34 & 0.08 & 0.24 & -- & -- & -- & 0.17 & 0.08 & 0.07 & -- & -- & -- \\ 
\end{longtable} 
\tablefoot{Abundances of elements derived on the scale of 
$\rm log \epsilon (X) = \rm log ( N (X) N (H )^{-1} ) + 12.0$, and relative 
to the solar [$X$] abundances based on the solar composition of 
\citet{Asplund2009,Scott2015a,Scott2015b}, and \citet{Grevesse2015}.}

\newpage

%-------------------------------------------------------------
\small
\begin{longtable}{llccccccccccccccc}
\caption{Abundances of elements for candidates for non-pulsating stars.}
\label{tab:chem_sum2} \\
\hline\hline
 &  & CAN-11 & & & CAN-13 & & & CAN-14 & & & CAN-16 & & & CAN-17 & & \\
Z & Element & $\rm log \epsilon (X)$ & $\pm 1\sigma$ & [$X$] & $\rm log \epsilon (X)$ & $\pm 1\sigma$ & [$X$] & $\rm log \epsilon (X)$ & $\pm 1\sigma$ & [$X$] & $\rm log \epsilon (X)$ & $\pm 1\sigma$ & [$X$] & $\rm log \epsilon (X)$ & $\pm 1\sigma$ & [$X$] \\
\hline
\endfirsthead
\caption{continued.}\\
\hline\hline
 &  & CAN-11 & & & CAN-13 & & & CAN-14 & & & CAN-16 & & & CAN-17 & & \\
Z & Element & $\rm log \epsilon (X)$ & $\pm 1\sigma$ & [$X$] & $\rm log \epsilon (X)$ & $\pm 1\sigma$ & [$X$] & $\rm log \epsilon (X)$ & $\pm 1\sigma$ & [$X$] & $\rm log \epsilon (X)$ & $\pm 1\sigma$ & [$X$] & $\rm log \epsilon (X)$ & $\pm 1\sigma$ & [$X$] \\
\hline
\endhead
\hline
\endfoot
6  & C  & 7.91 & 0.02 & -0.52 & 8.05 & 0.07 & -0.38 & 7.90 & 0.04 & -0.53 & 8.16 & 0.06 & -0.27 & 8.09 & 0.06 & -0.34 \\
7  & N  & -- & -- & -- & -- & -- & -- & -- & -- & -- & -- & -- & -- & -- & -- & -- \\
8  & O  & 8.64 & 0.06 & -0.05 & -- & -- & -- & 8.53 & 0.11 & -0.16 & -- & -- & -- & 8.63 & 0.08 & -0.06 \\
11 & Na & 6.08 & 0.03 & -0.13 & 5.98 & 0.04 & -0.23 & 6.06 & 0.03 & -0.15 & 6.40 & 0.07 & 0.19 & 6.15 & 0.06 & -0.06 \\
12 & Mg & 7.36 & 0.03 & -0.23 & 7.49 & 0.08 & -0.10 & 7.38 & 0.06 & -0.21 & 7.76 & 0.05 & 0.17 & 7.54 & 0.05 & -0.05 \\
13 & Al & -- & -- & -- & 6.71 & 0.07 & 0.28 & 6.12 & 0.05 & -0.31 & -- & -- & -- & -- & -- & -- \\
14 & Si & 7.37 & 0.03 & -0.14 & 7.29 & 0.04 & -0.22 & 7.31 & 0.03 & -0.20 & 7.59 & 0.08 & 0.08 & 7.38 & 0.05 & -0.13 \\
16 & S  & 6.95 & 0.03 & -0.17 & 7.00 & 0.03 & -0.12 & 6.81 & 0.03 & -0.31 & 7.16 & 0.05 & 0.04 & 6.89 & 0.04 & -0.23 \\
20 & Ca & 6.10 & 0.02 & -0.22 & 6.13 & 0.03 & -0.19 & 6.14 & 0.03 & -0.18 & 6.30 & 0.04 & -0.02 & 6.16 & 0.04 & -0.16 \\
21 & Sc & 2.74 & 0.03 & -0.42 & 2.66 & 0.03 & -0.50 & 2.66 & 0.03 & -0.50 & 2.94 & 0.06 & -0.22 & 2.68 & 0.04 & -0.48 \\
22 & Ti & 4.62 & 0.02 & -0.31 & 4.60 & 0.03 & -0.33 & 4.55 & 0.03 & -0.38 & 4.71 & 0.05 & -0.22 & 4.54 & 0.04 & -0.39 \\
23 & V  & 3.72 & 0.05 & -0.17 & 3.41 & 0.07 & -0.48 & 3.55 & 0.06 & -0.34 & 3.85 & 0.14 & -0.04 & 3.46 & 0.13 & -0.43 \\
24 & Cr & 5.39 & 0.03 & -0.23 & 5.37 & 0.04 & -0.25 & 5.37 & 0.04 & -0.25 & 5.64 & 0.05 & 0.02 & 5.38 & 0.05 & -0.24 \\
25 & Mn & 4.91 & 0.03 & -0.51 & 4.86 & 0.04 & -0.56 & 4.80 & 0.04 & -0.62 & 4.92 & 0.08 & -0.50 & 4.74 & 0.07 & -0.68 \\
26 & Fe & 7.15 & 0.02 & -0.32 & 7.16 & 0.03 & -0.31 & 7.16 & 0.03 & -0.31 & 7.38 & 0.04 & -0.09 & 7.16 & 0.03 & -0.31 \\
27 & Co & 4.46 & 0.22 & -0.47 & 4.37 & 0.10 & -0.56 & 4.37 & 0.06 & -0.56 & -- & -- & -- & -- & -- & -- \\
28 & Ni & 5.76 & 0.03 & -0.44 & 5.72 & 0.04 & -0.48 & 5.73 & 0.03 & -0.47 & 5.88 & 0.08 & -0.32 & 5.78 & 0.06 & -0.42 \\
29 & Cu & 3.96 & 0.06 & -0.22 & 3.52 & 0.06 & -0.66 & 3.52 & 0.03 & -0.66 & 3.83 & 0.17 & -0.35 & -- & -- & -- \\
30 & Zn & 4.03 & 0.05 & -0.53 & 4.12 & 0.04 & -0.44 & 3.81 & 0.03 & -0.75 & -- & -- & -- & 4.32 & 0.04 & -0.24 \\
38 & Sr & -- & -- & -- & 2.53 & 0.07 & -0.30 & 2.51 & 0.04 & -0.32 & -- & -- & -- & 2.69 & 0.09 & -0.14 \\
39 & Y  & 1.89 & 0.03 & -0.32 & 1.90 & 0.03 & -0.31 & 1.88 & 0.03 & -0.33 & 2.00 & 0.06 & -0.21 & 1.84 & 0.04 & -0.37 \\
40 & Zr & 2.45 & 0.05 & -0.14 & 2.39 & 0.06 & -0.20 & 2.37 & 0.05 & -0.22 & 2.42 & 0.12 & -0.17 & 2.30 & 0.08 & -0.29 \\
56 & Ba & 2.84 & 0.08 & 0.59 & 2.76 & 0.07 & 0.51 & 2.67 & 0.05 & 0.42 & 3.02 & 0.08 & 0.77 & 2.81 & 0.03 & 0.56 \\
57 & La & 1.41 & 0.02 & 0.30 & 1.28 & 0.04 & 0.17 & 1.27 & 0.03 & 0.16 & 1.39 & 0.14 & 0.28 & 1.29 & 0.05 & 0.18 \\
58 & Ce & 1.69 & 0.03 & 0.11 & 1.52 & 0.05 & -0.06 & 1.58 & 0.04 & 0.00 & 1.88 & 0.09 & 0.30 & 1.47 & 0.06 & -0.11 \\
59 & Pr & 0.83 & 0.05 & 0.11 & 0.66 & 0.07 & -0.06 & 0.78 & 0.05 & 0.06 & -- & -- & -- & 0.81 & 0.09 & 0.09 \\
60 & Nd & 1.50 & 0.04 & 0.08 & 1.44 & 0.05 & 0.02 & 1.48 & 0.03 & 0.06 & 1.80 & 0.13 & 0.38 & 1.38 & 0.06 & -0.04 \\
62 & Sm & 0.85 & 0.05 & -0.10 & 0.81 & 0.06 & -0.14 & 0.86 & 0.03 & -0.09 & 1.09 & 0.15 & 0.14 & 0.69 & 0.07 & -0.26 \\
63 & Eu & 0.84 & 0.02 & 0.32 & 0.89 & 0.05 & 0.37 & 0.92 & 0.04 & 0.40 & 1.36 & 0.13 & 0.84 & 0.89 & 0.08 & 0.37 \\
64 & Gd & 0.98 & 0.18 & -0.10 & 1.06 & 0.10 & -0.02 & 0.92 & 0.09 & -0.16 & -- & -- & -- & -- & -- & -- \\
65 & Tb & -- & -- & -- & -- & -- & -- & 0.13 & 0.14 & -0.18 & -- & -- & -- & -- & -- & -- \\
66 & Dy & -- & -- & -- & 0.73 & 0.14 & -0.37 & 1.06 & 0.11 & -0.04 & -- & -- & -- & 0.84 & 0.18 & -0.26 \\
68 & Er & -- & -- & -- & -- & -- & -- & -- & -- & -- & -- & -- & -- & -- & -- & -- \\
71 & Lu & 0.50 & 0.06 & 0.40 & -- & -- & -- & 0.33 & 0.03 & 0.23 & -- & -- & -- & 0.68 & 0.09 & 0.58 \\
\end{longtable} 
\tablefoot{Abundances of elements derived on the scale of 
$\rm log \epsilon (X) = \rm log ( N (X) N (H )^{-1} ) + 12.0$, and relative 
to the solar [$X$] abundances based on the solar composition of 
\citet{Asplund2009,Scott2015a,Scott2015b}, and \citet{Grevesse2015}.}

\end{landscape}

\FloatBarrier 
\twocolumn

\end{appendix}
\end{document}